%% file: paper.tex
\documentclass[sigplan,twocolumn,nonacm]{acmart}

\renewcommand\footnotetextcopyrightpermission[1]{}
\usepackage{pdflscape}
\usepackage{booktabs}
\usepackage{graphicx}
\usepackage{amsmath}
\usepackage[ruled,linesnumbered,vlined]{algorithm2e}
\usepackage{cleveref}
\crefname{algocf}{alg.}{algs.}
\Crefname{algocf}{Algorithm}{Algorithms}
\usepackage{tabularx}
\usepackage{enumitem}
\usepackage{booktabs}
\usepackage{multirow}
\usepackage{xcolor}
\usepackage{colortbl}

\usepackage{amsmath}
\usepackage{subcaption}
\usepackage{pifont}  

\usepackage{glossaries}

\newacronym{dnn}{DNN}{Deep Neural Network}
\newacronym{ml}{ML}{Machine Learning}
\newacronym{lstm}{LSTM}{Long Short-Term Memory}
\newacronym{sm}{SM}{Streaming Multiprocessor}
\newacronym{cuda}{CUDA}{Compute Unified Device Architecture}
\newacronym{opencl}{OpenCL}{Open Computing Language}
\newacronym{gpu}{GPU}{Graphics Processing Unit}
\newacronym{iot}{IoT}{Internet of Things}
\newacronym{slo}{SLO}{Service Level Objective}
\newacronym{tcn}{TCN}{Traffic Control Network}
\newacronym{yolo}{YOLO}{You Only Live Once}
\newacronym{ssd}{SSD}{Single Shot Multibox Detector}
\newacronym{ocr}{OCR}{Optical Character Recognition}
\newacronym{wrr}{WRR}{Weighted Round Robin}
\newacronym{cnn}{CNN}{Convolutional Neural Network}
\newacronym{cpu}{CPU}{Central Processing Unit}
\newacronym{cudnn}{cuDNN}{CUDA Deep Neural Network library}
\newacronym{dram}{DRAM}{Dynamic Random-Access Memory}
\newacronym{mps}{MPS}{Multi-Process Service}
\newacronym{mig}{MIG}{Multi-Instance GPU}
\newacronym{ai}{AI}{Artificial Intelligence}

\makeglossaries

\newcommand{\sysname}{\textsc{Roomie}}

\usepackage[all=normal,wordspacing]{savetrees}

\newcommand{\ie}{\emph{i.e.}\xspace}
\newcommand{\eg}{\emph{e.g.}\xspace}
\newcommand{\etal}{\emph{et al.}\xspace}

\renewcommand{\paragraph}[1]{\vspace*{0.05in}\noindent\textbf{#1}}

\title[\sysname: Interference-Aware Colocation for Efficient Model Serving]{\sysname: Interference-Aware Colocation\\for Efficient Model Serving}

\author{Youssouph Faye$^1$, Francescomaria Faticanti$^2$, Shubham Jain$^3$, Francesco Bronzino$^4$\\
{\small $^1$Université Savoie Mont Blanc, LISTC \quad $^2$Inria \quad $^3$Stony Brook University \quad $^4$ENS Lyon and Institut Universitaire de France}\\
}

\renewcommand{\shortauthors}{Y. Faye et al.}

\begin{document}

\input{sections/abstract}

\maketitle

\input{sections/introduction}

\input{sections/motivation}
\input{sections/kernel_interference_aware_scheduling}

\input{sections/implementation}

\input{sections/evaluation}
\input{sections/related}
\label{lastpage}\input{sections/conclusion}

\microtypesetup{protrusion=false}

\bibliographystyle{ACM-Reference-Format}
\bibliography{paper}
\vfill
\pagebreak

\end{document}

%% file: sections/abstract.tex
\begin{abstract}
As demand for \gls{dnn} inference grows, GPU capacity is increasingly oversubscribed, forcing
operators to colocate multiple models on the same device in both cloud and edge
deployments. Whether colocation succeeds or violates \gls{slo}s depends on the temporal overlap of kernels from concurrently executing models---an
effect that existing serving systems either ignore or approximate using
aggregate resource profiles that fail to capture temporal dynamics. This paper
presents {\sysname}, a model serving orchestration architecture that predicts
and avoids kernel-level interference between colocated DNNs. {\sysname}
decouples offline kernel profiling from online interference prediction.
It uses profiling only to extract per-kernel resource configurations, and predicts
interference with an occupancy-based analytical model immune to
profiler-induced timing distortion. 
A pairwise greedy heuristic then approximates multi-model interference in
polynomial rather than exponential time, and an online placement algorithm then uses
these estimates to assign each incoming model to the GPU that minimizes predicted
slowdown. Our experimental evaluation compares \sysname{} against
state-of-the-art solutions across both cloud-grade server clusters and embedded
edge devices, demonstrating that \sysname{} reduces \gls{slo} violations (\ie,
inference latency) by up to $3\times$, while maintaining comparable, and in many
cases superior, goodput relative to existing approaches.
\end{abstract}

%% file: sections/introduction.tex
\section{Introduction}\label{sec:intro}

\gls{ml} inference serving has become a foundational task for a variety of
domains, as organizations increasingly deploy \gls{ml} models with
applications spanning from computer vision to natural language
processing~\cite{taye2023understanding}. Unfortunately, the growing demand for
\gls{ml} inference requests is now outpacing hardware availability, creating a
critical resource gap that forces organizations to maximize utilization of
existing computational infrastructure. Operators are increasingly forced to
colocate multiple models on the same GPU, both in virtualized cloud clusters for cost
efficiency and on edge devices such as Nvidia Jetson
modules~\cite{nvidia2025jetson_modules}, where dedicated per-model hardware is
economically infeasible. In both settings, two models sharing a GPU can
interfere severely enough to violate their
\gls{slo}s~\cite{ahmad2024proteus,olston2017tensorflowserving,shubha2024usher,francisco2021infaas,mendoza2021interference}.
The placement decision, \ie, which models go on which GPU, determines
whether the system meets its latency targets or collapses under load. Existing
scalable serving stacks deployed in cloud platforms such as AWS
SageMaker~\cite{aws2017sagemaker} or Google Cloud AI~\cite{google2025cloud_ai}
typically rely on offline performance profiles to guide
placement~\cite{2017clipper,olston2017tensorflowserving,gujarati2020servingdnnslikeclockwork}.
These profiles capture latency or throughput under isolated conditions and
therefore yield limited benefit when models actually run concurrently and
interfere with one another.

Existing inference serving
systems~\cite{2017clipper,olston2017tensorflowserving,gujarati2020servingdnnslikeclockwork,francisco2021infaas}
do not adequately handle this colocation decision. Frameworks such as TensorFlow
Serving~\cite{olston2017tensorflowserving} and INFaaS~\cite{francisco2021infaas}
replicate models without modeling interference at all, treating GPUs as
interchangeable execution slots. Usher~\cite{shubha2024usher}, the closest prior
work, recognizes the problem and profiles models at the kernel level to
characterize their resource demands. It then pairs compute-heavy with memory-heavy
workloads on the assumption that complementary resource profiles imply
complementary execution. This assumption holds only at the aggregate level:
while inspecting kernels in isolation reveals their individual resource
footprints, it does not capture how those kernels overlap when
two models execute concurrently. Two models can have orthogonal resource
profiles on paper yet still collide repeatedly during execution, because their
kernel sequences happen to launch compute-bound stages at the same moments.
Conversely, two models with similar footprints may interleave gracefully if
their busy phases are offset. As we show in Section~\ref{sec:motivation}, this
gap between resource-level and execution-level reasoning causes Usher's
heuristic to produce colocations that violate SLOs even when its own resource
analysis predicts a good fit. Aggregate resource profiles miss what actually
matters: \emph{when} kernels from two models overlap in time, forcing sequential
execution and increasing latency.

Building a placement strategy on this insight requires solving two challenges.
\textbf{Challenge 1: profiling distortion.} Kernel-level profilers such as
Nsight and Torch Profiler are the only practical way to extract
per-kernel resource configurations on real hardware, but their instrumentation
inflates measured kernel durations by an average of 1.4$\times$ and up to
2.7$\times$ for workloads such as SSD on Nvidia A100 GPUs. Using raw profiled
durations as ground truth would therefore produce systematically inflated
interference predictions. \textbf{Challenge 2: combinatorial explosion.}
Interference between two colocated models depends on the \emph{alignment} of
their kernel sequences, and since models start asynchronously, every kernel can
in principle overlap with every other. Even ten models averaging fifty kernels
each yield 112{,}500 pairwise alignment scenarios, resulting in the search space
growing exponentially with the number of colocated models.

In this paper, we present {\sysname}, a model serving orchestration architecture
that tackles both challenges. To address Challenge~1, {\sysname} separates
profiling from prediction: profilers are used offline only to extract per-kernel
resource configurations (register usage, shared memory, and thread-block
dimensions), while latency predictions come from an occupancy-based analytical
model that is immune to profiler-induced timing distortion. To address
Challenge~2, {\sysname} reduces the alignment search space with a pairwise
greedy heuristic that evaluates a sampled subset of starting indices between
each pair of models and aggregates the result using a robust median estimator,
capturing the temporal dynamics of interference without exhaustive enumeration.
An online placement algorithm then uses these interference estimates to assign
incoming models to the GPU that minimizes the average predicted performance drop
across colocated workloads, subject to a tunable degradation bound.

We evaluate {\sysname} on a 12-GPU Nvidia A100 cluster and a 12-device Jetson
Xavier edge deployment, against INFaaS~\cite{francisco2021infaas} and
Usher~\cite{shubha2024usher} on both real~\cite{twitterStreamTrace2020} and
synthetic workloads. By accurately modeling interference and guiding colocation
decisions, {\sysname} sustains responsiveness under heavy workloads, reducing
SLO violations to 3$\times$ lower than INFaaS~\cite{francisco2021infaas} and
2$\times$ lower than Usher~\cite{shubha2024usher} in cloud clusters. On edge
devices, {\sysname} achieves similar or superior performance, keeping violations
far below competing baselines even under tight resource constraints.
Placement-accuracy experiments confirm that {\sysname} matches optimum placement
in 90\% of randomized trials, while decision latency stays below one second even
for fifteen models on ten GPUs.

In summary, our contributions are:
\begin{itemize}[leftmargin=*,noitemsep,topsep=0pt]
\item A \emph{kernel-aware profiling and interference estimation} framework that
extracts per-kernel resource configurations offline and uses them to efficiently
approximate interference impact via a pairwise greedy heuristic, avoiding
exhaustive enumeration of kernel overlap scenarios
(\Cref{sec:kernel_interference_aware_scheduling}).
\item {\sysname}, a model serving system that integrates the above offline
profiling pipeline with an \emph{online placement algorithm} that dynamically
orchestrates model placement to minimize SLO violations while maintaining
goodput on resource-constrained deployments (\Cref{sec:implementation}).
\item A thorough evaluation on both a 12-GPU Nvidia A100 cloud cluster and a
12-device Nvidia Jetson Xavier edge deployment, demonstrating consistent
superiority over state-of-the-art baselines across real and synthetic workloads
(\Cref{sec:evaluation}).
\end{itemize}

%% file: sections/motivation.tex
\section{Motivation}\label{sec:motivation}
As demand for real-time inference grows, organizations must maximize the
utilization of existing infrastructure. Unfortunately, existing orchestration
solutions rely on simplistic heuristics for model placement and colocation,
typically referencing offline profiling data or prioritizing devices with the
most available memory. We first argue that effective colocation requires
reasoning about kernel execution patterns rather than aggregate resource
profiles, and that the temporal alignment of those kernels---not just their
resource footprints---governs interference. We then show how a representative
state-of-the-art system~\cite{shubha2024usher} fails precisely because it
overlooks this temporal dimension. Finally, we describe the practical challenges
of building accurate performance models at kernel granularity.

\subsection{Kernel Execution Patterns Govern Interference}\label{sec:motivation:kernels}

\begin{figure}[t!]
	\centering
	\begin{subfigure}[b]{0.46\linewidth}
		\centering
		\includegraphics[width=\linewidth]{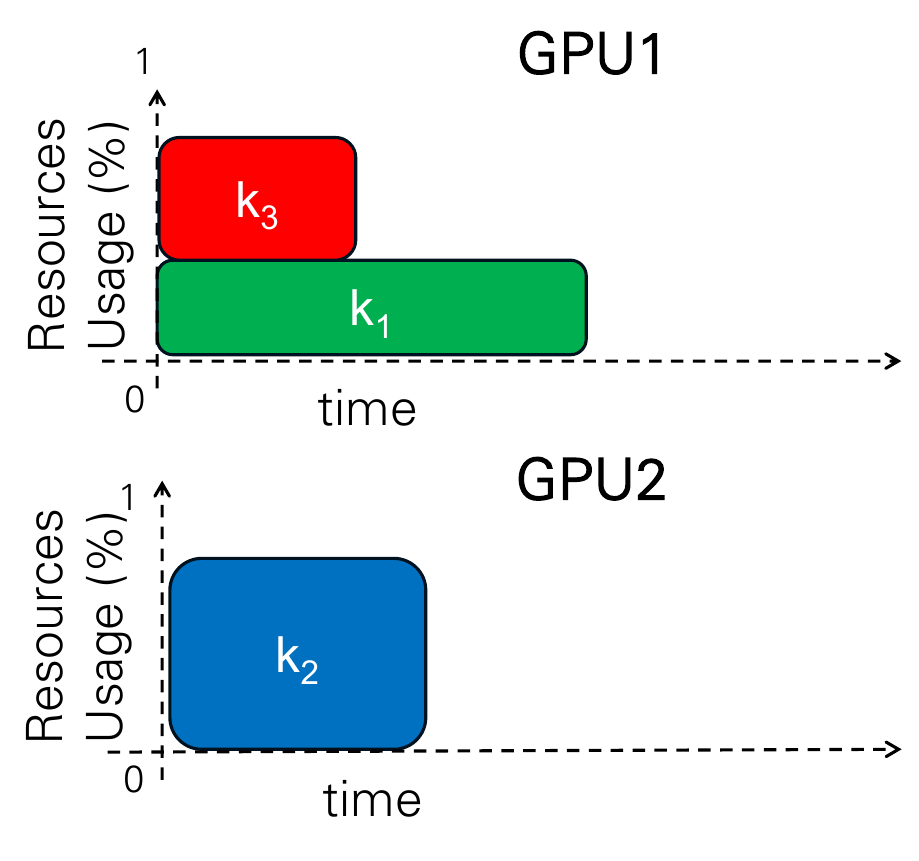}
		\subcaption{Proper colocation: $k_1$ and $k_3$ coexist on GPU1 while $k_2$ executes on GPU2, maximizing utilization without resource contention.}
		\label{fig:colocation_a}
	\end{subfigure}
	\hfill
	\begin{subfigure}[b]{0.46\linewidth}
		\centering
		\includegraphics[width=\linewidth]{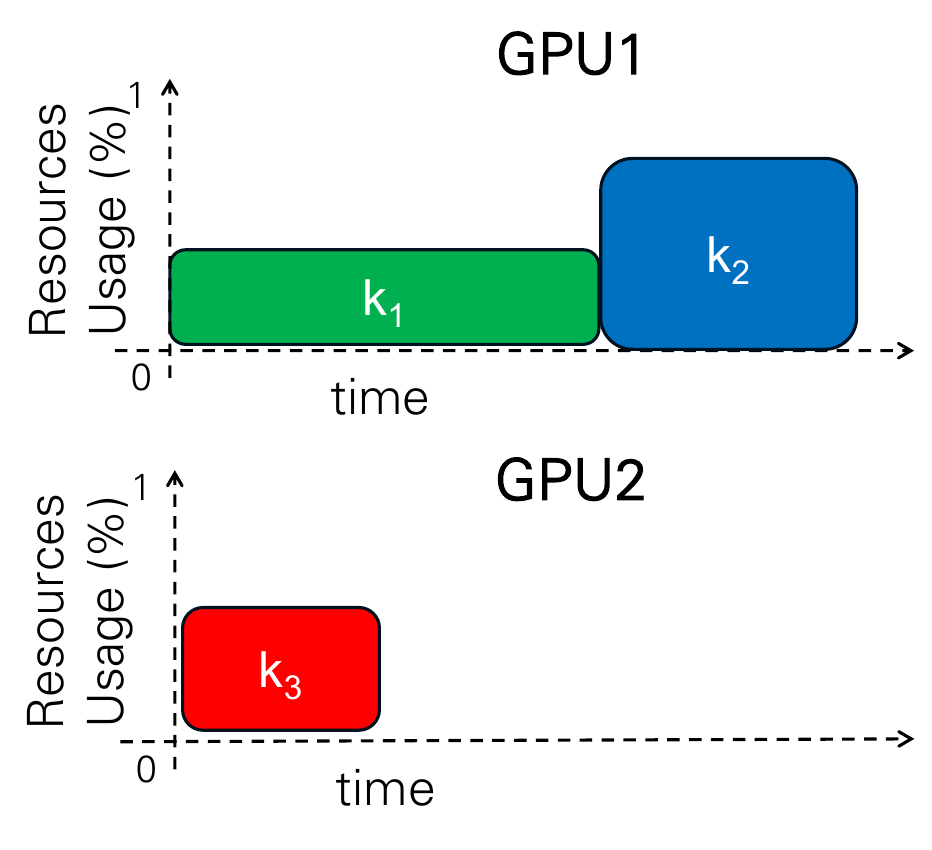}
		\subcaption{When combined resource demands of $k_1$ and $k_2$ exceed GPU capacity, kernels are executed sequentially, eliminating parallelization benefits.}
		\label{fig:colocation_b}
	\end{subfigure}
	\caption{Kernel execution patterns under different colocation strategies demonstrate the impact of placement on execution time.}
	\label{fig:colocation}
\end{figure}

\glspl{dnn} perform inference by executing a sequence of \emph{kernels}---the
fundamental unit of GPU computation---each responsible for a specific low-level
operation with its own resource demands and execution duration. Kernels, rather
than high-level model architectures or aggregate memory statistics, constitute
the true computational footprint on the GPU and determine performance under
colocation.

When two kernels from different models execute concurrently and their combined
resource demands stay within GPU capacity, they run in parallel and each
completes at roughly its isolated latency (Figure~\ref{fig:colocation_a}). When
those demands exceed capacity, the hardware must serialize execution: one kernel
waits while the other progresses, and inference latency rises accordingly
(Figure~\ref{fig:colocation_b}). Crucially, whether two colocated models fall
into the first or second regime is not a property of their average resource
consumption, but of \emph{when} their resource-intensive kernels happen to
execute. Two models with orthogonal resource profiles can still collide
repeatedly if their heaviest kernels launch at the same moments; conversely, two
models with similar footprints may interleave gracefully if their busy phases
are offset. Effective colocation therefore requires reasoning about the temporal
alignment of kernel sequences, not just their aggregate demands---an aspect
invisible to resource-only models.

\subsection{Why Resource-Complementarity Heuristics Fail}\label{sec:motivation:usher}

\begin{table}[t]
\centering
\small
\caption{Inference latency comparison for isolated execution and colocation. The SLO threshold is set to $2\times$ the isolated latency~\cite{li2023alpaserve, choi2022serving, shubha2024usher}.}
\label{tab:goodput_comparison}

\begin{subtable}[t]{\columnwidth}
\centering
\caption{Isolated execution}
\label{tab:isolated}
\setlength{\tabcolsep}{8pt}
\begin{tabular}{@{}lrr@{}}
\toprule
{\bf Model} & {\bf Avg. latency (ms)} & {\bf SLO threshold (ms)} \\
\midrule
AlexNet   & 0.50  & 1.00   \\
GoogLeNet & 3.29  & 6.58   \\
SSD       & 66.67 & 133.34 \\
\bottomrule
\end{tabular}
\end{subtable}

\vspace{0.3cm}

\begin{subtable}[t!]{\columnwidth}
\centering
\caption{Colocation on two GPUs}
\label{tab:colocation}
\setlength{\tabcolsep}{6pt}
\begin{tabular}{@{}lllrc@{}}
\toprule
{\bf System} & {\bf GPU} & {\bf Model} & {\bf Avg. lat. (ms)} & {\bf Violates SLO} \\
\midrule
\multirow{3}{*}{Usher}
  & \multirow{2}{*}{GPU1} & AlexNet   & 1.03   & \ding{51} \\
  &                       & SSD       & 333.33 & \ding{51} \\
\cmidrule(l){2-5}
  & GPU2                  & GoogLeNet & 3.29   & \ding{55} \\
\midrule
\multirow{3}{*}{{\sysname}}
  & \multirow{2}{*}{GPU1} & GoogLeNet & 3.33   & \ding{55} \\
  &                       & SSD       & 200.00 & \ding{51} \\
\cmidrule(l){2-5}
  & GPU2                  & AlexNet   & 0.50   & \ding{55} \\
\bottomrule
\end{tabular}
\end{subtable}
\end{table}

The previous section argues that systems reasoning about colocation in
resource-profile terms will systematically misidentify good pairings. We
illustrate this concretely with Usher~\cite{shubha2024usher}, the most
prominent recent attempt to optimize colocation through kernel-level resource
analysis. Usher analyzes low-level GPU kernels to estimate resource demands and
couples compute-intensive models with memory-intensive ones to maximize
utilization. This logic is appealing but assumes that contrasting resource
profiles imply non-overlapping execution, an assumption that holds only when
kernel timing happens to be favorable.

To make the failure mode visible, we evaluate three models that span Usher's
decision space on two Nvidia Jetson AGX Xavier devices: AlexNet
(compute-heavy), SSD (memory-heavy), and GoogLeNet (balanced).
\Cref{tab:isolated} reports their isolated latencies and the corresponding SLO
thresholds, defined as twice the isolated inference
time~\cite{li2023alpaserve,choi2022serving,shubha2024usher}.

Following its heuristic, Usher colocates AlexNet (compute-heavy) with SSD
(memory-heavy) on GPU1 and leaves GoogLeNet to run alone on GPU2. As shown in
\Cref{tab:colocation}, this configuration fails to deliver the expected
benefits: AlexNet's latency grows from 0.5\,ms to 1.03\,ms (2.06$\times$) and
SSD's from 66.67\,ms to 333.33\,ms (5$\times$), with both violating their SLO
thresholds. An alternative placement that takes kernel timing into account
yields markedly better results. Colocating GoogLeNet with SSD on GPU1 and
running AlexNet alone on GPU2 keeps GoogLeNet within 1\% of its baseline
(3.33\,ms vs.\ 3.29\,ms), reduces SSD to 200\,ms (40\% better than under Usher),
and lets AlexNet run at its isolated latency with zero interference. The pairing
Usher's heuristic identifies as ideal turns out to be the worst of the available
options, while a placement its heuristic cannot recognize keeps two of three
models within SLO. The cause is precisely what
Section~\ref{sec:motivation:kernels} predicts: aggregate resource profiles are
not a sufficient signal about \emph{when} kernels contend, and any heuristic
built on them will miss colocations governed by temporal alignment rather than
resource arithmetic.

\subsection{Challenges in Modeling Kernel-Level
Interference}\label{sec:motivation:challenges}

A kernel-aware approach must overcome two obstacles that have so far prevented
prior work from exploiting kernel timing directly.

\paragraph{Profiling distortion.} Tools such as
Nsight-Systems~\cite{nsight_systems} and Torch Profiler~\cite{torch_profiler}
are the only practical way to extract per-kernel resource configurations and
execution characteristics on real hardware. However, the instrumentation and
callbacks they introduce add latency to kernel execution: in our measurements,
recorded durations exceed true inference time by an average of 1.4$\times$,
reaching 2.7$\times$ for workloads such as SSD on Nvidia A100 GPUs. Using raw
profiler measurements to estimate colocation performance would therefore yield
systematically inflated latency predictions and incorrect placement decisions.
Profiling remains necessary to capture kernel structure on specific hardware,
but performance prediction must be decoupled from raw profiled durations.

\paragraph{Combinatorial explosion of kernel alignments.} Even if profiling were
free, exhaustively measuring every alignment scenario is infeasible.
Interference arises dynamically as kernels from different models overlap, and
since models do not begin inference synchronously, each can start at any point
in its kernel sequence. With just 10 models averaging 50 kernels each, the
number of pairwise overlap scenarios already reaches 112{,}500---and the space
grows exponentially when more than two models share a GPU. An accurate
interference model must therefore estimate alignment effects without enumerating
them.

These two obstacles jointly motivate our approach. Profiling is used offline to
extract kernel structure and resource configurations, while a dedicated
occupancy-based model uses those characteristics to predict interference without
relying on raw profiled durations; a pairwise greedy heuristic then approximates
multi-model alignment effects in polynomial rather than exponential time. We
describe this design next.

%% file: sections/kernel_interference_aware_scheduling.tex
\section{Kernel Interference-Aware Scheduling}\label{sec:kernel_interference_aware_scheduling}

To address the challenges of profiling overhead and combinatorial complexity, we
introduce a kernel-level profiling and interference estimation
strategy that balances precision with scalability. We address
Challenge 1 by separating profiling (which captures hardware-specific execution
characteristics) from performance estimation (which models resource occupation
and execution behavior). We address Challenge 2 by analyzing execution traces
in isolation and modeling representative overlap scenarios, enabling informed
deployment decisions without exhaustively evaluating all possible alignments. In
this section we explain the main concepts behind our approach. First, we formalize
the notion of interference and describe a method for estimating it. Next, we
present an algorithm that efficiently calculates interference among various
models. Finally, we detail a placement algorithm that utilizes the interference
estimation to efficiently allocate incoming models to available GPUs.

\subsection{Kernel Execution and Occupancy Modeling}\label{subsec:kernel_execution_and_occupancy_modeling}

DNN inference on Nvidia GPUs proceeds as a sequence of CUDA kernels, each responsible for one or more model layers (\eg, convolution, activation, pooling). How well a kernel utilizes GPU resources determines its performance and, under colocation, its susceptibility to interference. We characterize this through theoretical occupancy: the fraction of warp capacity actively used by a kernel on a given streaming multiprocessor (SM). 

CUDA threads are organized into warps — groups of 32 threads that execute instructions in lockstep — which are further grouped into blocks, the basic scheduling units on the GPU. The maximum number of blocks $b$ that can run concurrently on an SM is determined by the most restrictive of three resource constraints~\cite{lim2017autotuninggpukernelsstatic}: while the SM imposes a fixed upper bound on active warps $W$, each block also consumes registers and shared memory, both available in finite quantities per SM. If either of these is exhausted before the warp limit is reached — leaving the kernel \textit{register-limited} or \textit{shared-memory-limited} — the number of blocks that can be accommodated is reduced accordingly. Given $b$, and the number of warps per block $W_b$ fixed by the kernel's thread configuration, theoretical occupancy is:
\begin{equation}\label{eq:occupancy}
	o = \frac{W_b \cdot b}{W},
\end{equation}
where $W$ is the SM's maximum supported warp count. This ratio expresses how effectively a kernel uses the GPU under exclusive access, and serves as the baseline for estimating performance degradation when that exclusivity is broken by colocation.

\subsection{Interference and Adjusted Occupancy}\label{subsec:adjusted_occupancy}

Interference arises when the combined resource demand of concurrently scheduled kernels exceeds the GPU's capacity. Formally:
\begin{equation}\label{eq:interf}
	\sum_{k \in K} \varphi_k > \Phi,
\end{equation}
where $K$ is the set of active kernels, $\varphi_k$ is the resource usage of kernel $k$, and $\Phi$ represents the total GPU resource capacity. When this condition holds, the hardware cannot schedule all kernels at their intended occupancy, forcing some to run with fewer active warps or blocks than their standalone configuration would allow, increasing inference latency.

\paragraph{Adjusted occupancy}. We quantify this degradation by recomputing each kernel's occupancy under the reduced resource share it receives. The constrained number of blocks $\tilde{b}_k$ is derived from the same SM constraints as in~\Cref{subsec:kernel_execution_and_occupancy_modeling}, but applied to the resources remaining after accounting for co-scheduled kernels. With $\tilde{b}_k$, we can derive the new occupancy $\tilde{o}_k$, using the same formulation as in~\Cref{eq:occupancy}, from which we estimate the kernel's new execution time:
\begin{equation}
	\tilde{d}_k^e = d_k \cdot \frac{o_k}{\tilde{o}_k},
\end{equation}
where $d_k$ and $o_k$ represent the kernel's isolated execution time and theoretical occupancy, respectively. Execution time is thus inversely proportional to adjusted occupancy. If $\tilde{o}_k = o_k$, no inference occurs.

Since GPU scheduling policy under concurrent execution is not fixed~\cite{nvidia2025cuda}, we approximate it using simplified classical strategies (equal partitioning, priority-based, or first-come-first-served), which provide a practical basis for estimating $\tilde{b}_k$, without requiring any knowledge of the exact scheduler behavior.

\paragraph{Two-phase execution}. Assuming that two kernels are interfering, such an interference ends once the shortest-running kernel completes. Let $\Delta = \min_{k} \{\tilde{d}_k^e\}$, the remaining kernel transitions from reduced occupancy $\tilde{o}_k$ to full occupancy $o_k$. Its total adjusted duration is:
\begin{equation}
	\tilde{d}_k =
	\begin{cases}
		\tilde{d}_k^e & \text{if } \tilde{d}_k^e \leq \Delta \\
		\Delta + \left( \tilde{d}_k - \Delta \right) \cdot \frac{\tilde{o}_k}{o_k} & \text{otherwise.}
	\end{cases}
\end{equation}
The portion $\tilde{d}_k^e - \Delta$, originally computed under reduced occupancy, is scaled by $\frac{\tilde{o}_k}{o_k}$ to reflect the normal execution once interference ends.

\paragraph{Performance drop}. For a model $m$ with a sequence of $q$ kernels, we align the first kernels of each colocated model and simulate execution iteratively: when the active kernel from one model completes, it is replaced by the next kernel in the model's sequence. The total inference time under interference is 
\begin{equation*}
	\tilde{T}_m = \sum_{i=1}^{q} \tilde{d}_{k_i},
\end{equation*}
and the performance drop experienced by model $m$ due to interference is quantified by the relative increase in inference time. This is computed as:
\begin{equation}\label{eq:performance_drop}
	\mu_m = \frac{\sum_{i=1}^{q} (\tilde{d}_{k_i} - d_{k_i})}{\sum_{i=1}^{q} d_{k_i}} = \frac{\tilde{T}_m - T_m}{T_m}.
\end{equation}

\paragraph{Kernel alignment}. The analysis above assumes models start from their first kernel simultaneously, but in practice any kernel in one model's sequence may overlap with any kernel in the sequence of another model. Each such alignment yields a distinct performance drop, and the number of possible alignments among all the kernels of all the co-located models grows combinatorially with model count and sequence length, making exhaustive evaluation impractical. This motivates the heuristic introduced in~\Cref{greedy_algorithm}. 

\subsection{Greedy Algorithm for Estimating Model Interference}\label{greedy_algorithm}

Evaluating all possible kernel alignments across $N$ concurrent models requires constructing a Cartesian product of starting indices (one per model) leading to an exponential growth in the search space. We address this with a greedy pairwise heuristic that reduces the search space to the minimal meaningful subset: as demonstrated in~\Cref{sec:evaluation}, considering only pairwise interactions between models, rather than all $N$-way combinations, is sufficient to achieve significant reductions in SLO violations while keeping the estimation tractable. The pseudocode of the algorithm is presented in~\Cref{algo:kernel_interference_algorithm}.

\begin{algorithm}[t!]
	\caption{Greedy Estimation of Model Interference}
	\label{algo:kernel_interference_algorithm}
	\SetAlgoLined
	\SetKwProg{Fn}{Function}{}{end} 
	\Fn{performance\_drop($M$)}
	{
		\begin{small}
			Define $\mathcal{C}_{i,j}$ for each pair of models $m_i$, $m_j$

			Initialize performance drop $\mu \gets []$

			\ForEach{model $m_i \in M$}{
				Initialize $\tilde{T_{m_i}} \gets \sum d_k$

				\ForEach{model $m_j$ where $j \neq i$}{
					Initialize delay set $\mathcal{D}_{i,j} \gets []$

					\ForEach{pair $(s_i, s_j) \in \mathcal{C}_{i,j}$}{
						$k_i \gets s_i$, $k_j \gets s_j$, $\delta \gets 0$

						\While{$k_i < q_i$ and $k_j < q_j$}{
							\If{$\varphi_{k_i} + \varphi_{k_j} > \Phi$}{
								Compute additional duration:

								$\delta \gets \delta + d_{k_i} \cdot \frac{o_{k_i}}{o_{k_i} + o_{k_j}}$
							}
							Increment $k_i$, $k_j$
						}
						Append $\delta$ to $\mathcal{D}_{i,j}$
					}
					Compute overlap factor $\gamma_{i,j} \gets \max\left(\frac{q_i}{q_j}, 1\right)$

					Update $\tilde{T_{m_i}} \gets \tilde{T_{m_i}} + \gamma_{i,j} \cdot \text{median}(\mathcal{D}_{i,j})$
				}
				$\mu_{m_i} \gets \frac{\tilde{T}_{m_i} - T_{m_i}}{\tilde{T}_{m_i}}$

				Append $\mu_{m_i}$ to $\mu$
			}
			\Return{$\mu$}
		\end{small}
	}
\end{algorithm}

\paragraph{Pairwise interference}. For each pair $(m_i,m_j)$, the algorithm defines a set $\mathcal{C}_{i,j}$ of starting index pairs $(s_i,s_j)$ representing candidate alignment points (line 2). This set is obtained by starting from the Cartesian product of the sets of kernels' indices of the two models, and considering only the pairs of indexes that represent a feasible alignment and an interference starting point. For each $(s_i,s_j) \in \mathcal{C}_{i,j}$, the algorithm simulates kernel-by-kernel execution (lines 10-16): whenever the combined resource demand $\varphi_{k_i} + \varphi_{k_j} > \Phi$ (line 11), the delay added to kernel $k_{s_i}$ (\ie, the kernel identified by the starting point $s_i$) is $\delta_{k_{s_i}} = d_{k_{s_i}} \cdot o_{k_{s_i}}/(o_{k_{s_i}} + o_{k_{s_j}})$ (line 13). The total delay accumulated over the alignment is (line 17):
\begin{equation*}
	\Delta^{c_{i,j}} = \sum_{ s_i \leq t \leq q_i} \delta_{k_t}.
\end{equation*}
We take the median across all the alignments as the representative delay, chosen for its robustness to outlier configurations (line 20):
\begin{equation*}
\Delta_{i,j} = \text{median}\left(\left\{ \Delta^{c_{i,j}} \right\}_{\forall c_{i,j} \in \mathcal{C}_{i,j}}\right)
\end{equation*}

\begin{algorithm}[t]
	\caption{Model Placement Algorithm}
	\label{algo:model_placement}
	\SetAlgoLined
	\SetKwProg{Fn}{Function}{}{end} 
	\Fn{schedule($m_{arr}, G, \lambda$)}
	{
		\begin{small}
			Initialize $p \gets []$ be sequence of performance drops of new model $m_{arr}$.

			Initialize performance drop $\mathcal{P} \gets []$

			\ForEach{ GPU $g$ }{
			$M^g \gets M^g \cup \{m_{arr}\}$

			$\mathbf{\mu}^g \gets performance\_drop($M$)$

			\tcc{Ensure no variant has a performance drop beyond $\lambda$.}

			\If{ $\bar{\mu^g} < \lambda$ }{
			Append $\mu^g_{m_{arr}}$ to $p$

			Append $\mathbf{\mu}^g$ to $\mathcal{P}$
			}
			}



			Peak lowest average performance drop.

			$g* \gets \min \left\{ \bar{\mu}^g \right\}_{\forall \mathbf{\mu}^g \in \mathcal{P}}$

			\Return{$g^*$}
		\end{small}
	}
\end{algorithm}

\paragraph{Overlap scaling}. To account for the amount of time models $m_i$ and $m_j$interact during execution, we introduce a scaling factor $\gamma_{i,j} = \max\left(\frac{q_i}{q_j}, 1\right)$ that reflects their relative kernel sequence lengths (line 19). The estimated inference time of model $m_i$ under interference from all co-scheduled models is then (line 20):
\begin{equation*}
\tilde{T_{m_i}} = T_{m_i} + \sum_{\substack{j=1 \\ j \neq i}}^{N} \gamma_{i,j} \cdot \Delta_{i,j},
\end{equation*}
and the performance drop $\mu_{m_i}$ follows from~\Cref{eq:performance_drop} (lines 22 and 23).

\subsection{Efficient Placement}\label{sec:placement}

The performance drop estimated by~\Cref{algo:kernel_interference_algorithm} can be directly exploited to guide the placement of incoming models, whether triggered by a new deployment request or by the need to scale up and meet increased workload demand. \Cref{algo:model_placement} describes the proposed procedure to place a new model that needs to be deployed. When a new model $m_{arr}$ arrives, \Cref{algo:kernel_interference_algorithm} is applied to each candidate GPU $g$ (line 5) to estimate the average performance drop $\bar{\mu}^g$ across all models that would run on $g$, including $m_{arr}$. The threshold $\lambda$, taken as input (line 1), represents the maximum performance degradation (in percentage) that $m_{arr}$ is allowed to impose on already-running models. Among all GPUs satisfying $\bar{\mu}^g < \lambda$ (line 7), the one yielding the lowest $\bar{\mu}^g$ is selected as the deployment target (line 13); if none qualifies, the deployment is deferred. It is worth noting that the placement objective can be easily substituted for alternative, such as maximizing throughput or minimizing latency, without changing the structure of the algorithm.

%% file: sections/implementation.tex
\section{\sysname: System Design}\label{sec:implementation}

\begin{figure}[t]
    \centering
    \includegraphics[width=\columnwidth]{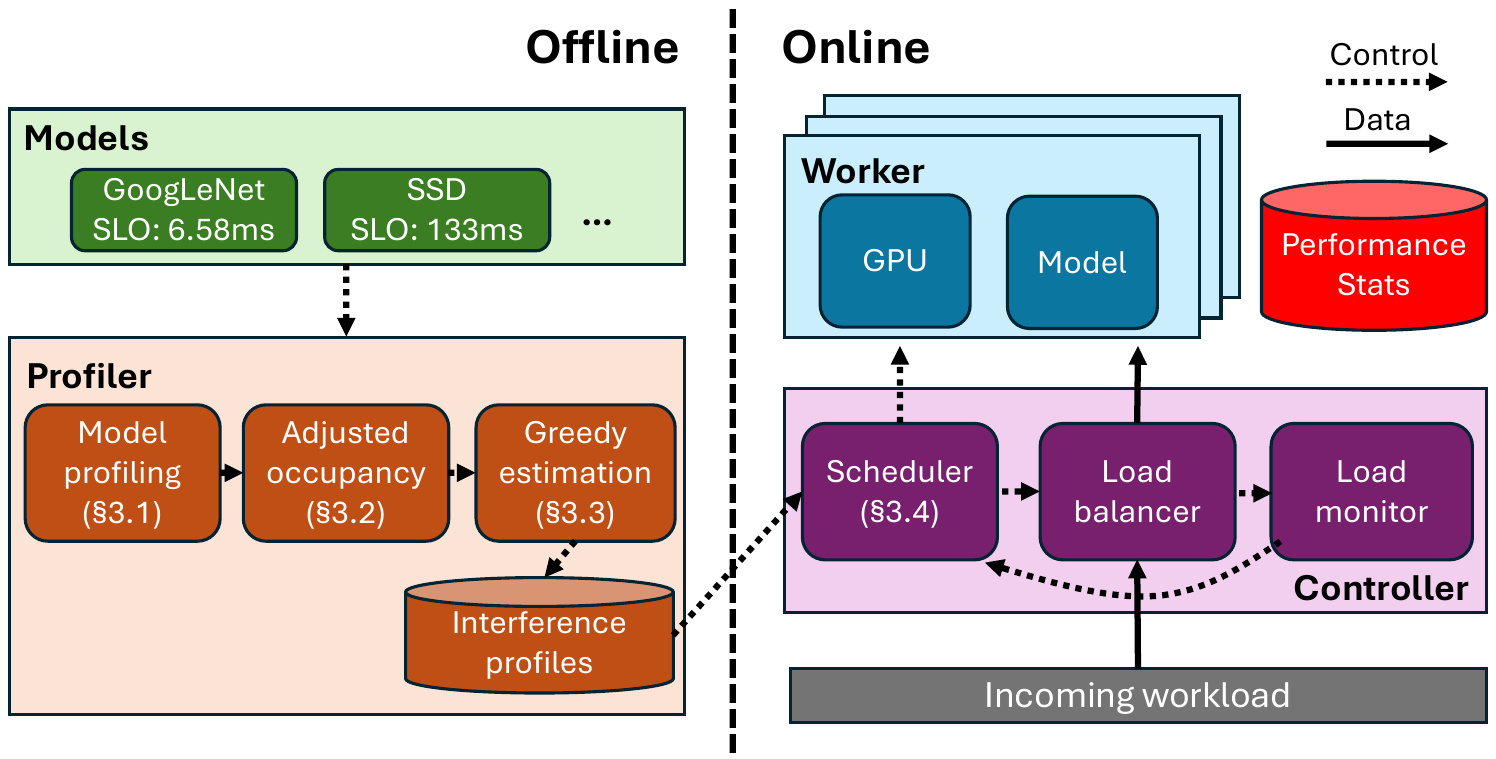}
    \caption{Overview of the {\sysname} architecture, showing the offline profiling pipeline and the online serving stack.}
    \label{fig:roomie}
\end{figure}

We describe the implementation of {\sysname}, a model serving system that
integrates the kernel-aware profiling and interference estimation into an online placement
pipeline. Figure~\ref{fig:roomie} illustrates the {\sysname} architecture, which
is organized into two phases. The \emph{offline phase} is carried out by a
\emph{Profiler} that executes the three stages introduced in
Section~\ref{sec:kernel_interference_aware_scheduling}, \ie, model profiling, adjusted
occupancy computation, and greedy interference estimation, to produce, for every
supported DNN, the interference profiles required for placement. The
\emph{online phase} is carried out by a \emph{Controller} that uses these
profiles to run the placement algorithm of \Cref{sec:placement},
deciding which GPU each model is deployed on, routing incoming queries to the
appropriate workers, and reacting to runtime performance feedback. {\sysname} is
implemented in approximately 15,000 lines of C++ and Python code: the client and
controller are written in C++ for performance, while the profiler and workers
use Python to leverage the PyTorch framework. Components communicate via
WebSocket for efficient asynchronous message exchange.

\paragraph{Profiler.}
The Profiler implements the offline pipeline that produces the interference
profiles consumed by the online scheduler. It executes three stages, mirroring
the methodology of Section~\ref{sec:kernel_interference_aware_scheduling}:
\emph{(i)}~\emph{model profiling}~(\Cref{subsec:kernel_execution_and_occupancy_modeling}), which
extracts the per-kernel resource configurations—register usage, shared memory,
and thread-block dimensions—from each DNN; \emph{(ii)}~\emph{adjusted occupancy}
computation~(\Cref{subsec:adjusted_occupancy}), which derives the resource-aware
occupancy of every kernel under contention; and \emph{(iii)}~\emph{greedy
estimation}~(\Cref{greedy_algorithm}), which aggregates these quantities into
pairwise interference profiles without exhaustively enumerating kernel
alignments. The first stage relies on platform-specific tooling: on Jetson
devices we use Nvidia Nsight-Compute~\cite{ncu2025}, while on A100 systems we
employ the PyTorch Profiler~\cite{pytorchProfiler2025}, since PyTorch Profiler
is unsupported in Jetson L4T containers and Nsight-Compute presented
compatibility issues on our A100 hardware. The resulting interference profiles
are persisted and made available to the online phase.

\paragraph{Controller.} The Controller orchestrates the online phase and is
composed of three cooperating components: a \emph{scheduler}, a \emph{load
balancer}, and a \emph{load monitor}. The scheduler runs the placement
algorithm of \Cref{sec:placement} to decide which GPU each model is
deployed on: for every newly arriving model or scaling event, it consults the
offline interference profiles and selects the GPU that minimizes the predicted
performance degradation across the resulting set of co-located models. The
chosen model-to-worker mapping is then handed to the load balancer, which
routes incoming inference queries to the appropriate workers. Finally, the
load monitor continuously observes per-worker performance statistics and feeds
them back to the scheduler, which uses them to detect violations of the
degradation budget and trigger re-placement or scaling decisions when needed.

\paragraph{Inference runtime.}
DNN inference is powered by PyTorch, using pretrained classification and
detection models from TorchVision. Inference workers are deployed in
containerized environments adapted to the underlying hardware: cloud systems
with Nvidia A100 \glspl{gpu} use the
\textit{pytorch:2.5.0-cuda12.1-cudnn9-runtime} container, while edge deployments
on Jetson Xavier devices use the ARM64-optimized
\textit{dustynv/l4t-pytorch:r35.4.1} image with native CUDA support. Workers handle incoming queries by assigning them to queues dedicated to each deployed model. Each DNN instance operates in its own CUDA Stream~\cite{nvidia2025cuda}, enabling multiple models to execute concurrently on the same device. For each model, the instance retrieves queries from its queue and assembles them into batches up to the batch size configured for that model before issuing an inference pass.

\paragraph{Runtime monitoring.}
GPU utilization is monitored using nvitop~\cite{nvitop} on A100 systems and jetson\_stats~\cite{jetson_stats} on Jetson devices, providing runtime visibility for stable operation under varying workloads.

\paragraph{Workload generation.} Finally, to evaluate the system under
controlled load, the client includes a lightweight traffic generator that issues
queries at configurable rates, allowing us to emulate the workload patterns used
in our evaluation. The client takes as input a trace and adapts it according to
the target workload characteristics (more details in Section~\ref{sec:setup}).


\section{Experimental Setup}\label{sec:setup}

\begin{table*}[t]
	\centering
	\small
	\caption{Server configuration used for experiments}
	\label{tab:serve_config}
	\begin{tabular}{@{}llc p{.2\textwidth} p{.28\textwidth}@{}}
		\toprule
		\textbf{Use Case} & \textbf{Model} & \textbf{\# Servers} & \textbf{CPU Configuration} & \textbf{GPU} \\
		\midrule
		\multirow{2}{*}{Cloud GPU Cluster} 
		& Apollo 6500 Gen10+ & 3 & \mbox{1$\times$ Intel Xeon, 32 cores} & $4\times$ Nvidia A100-SXM4-40GB, CC\textsuperscript{a}: 8.0 \\
		& DL360 Gen10+ & 2 & \mbox{2$\times$ Intel Xeon, 32 cores} & -- \\
		\midrule
		\multirow{2}{*}{Edge GPU Cluster} 
		& Nvidia Jetson AGX Xavier & 12 & \mbox{1$\times$ ARMv8, 8 cores} & Nvidia AGX Xavier, CC\textsuperscript{a}: 7.2 \\
		& HPE Proliant DL360 Gen10+ & 2 & \mbox{2$\times$ Intel Xeon, 16 cores} & -- \\
		\bottomrule
	\end{tabular}

	\vspace{0.1cm}
	{\footnotesize \textsuperscript{a}CC: Compute Capability}
\end{table*}

\begin{table*}[t]
	\centering
	\small
	\caption{Categorization of \gls{dnn} Models Used in Evaluation.
		Models marked with $^\dagger$ were used only in the cloud cluster evaluation;
		all others were included in both cluster and Jetson Xavier evaluation.}
	\label{tab:dnn-models}
	\begin{tabular}{@{}lp{0.8\linewidth}@{}}
		\toprule
		\textbf{Category}       & \textbf{Models}                                             \\
		\midrule
		Classification Models   &
		\texttt{alexnet}, \texttt{maxvit\_t}, \texttt{googlenet},
		\texttt{densenet201}, \texttt{mobilenet\_v3\_large},
		\texttt{squeezenet1\_1}, \texttt{shufflenet\_v2\_x2\_0},
		\texttt{inception\_v3}, \texttt{vgg19},
		\texttt{resnet152}$^\dagger$, \texttt{wide\_resnet101\_2}$^\dagger$,
		\texttt{resnext101\_32x8d}$^\dagger$, \texttt{efficientnet\_v2\_l}$^\dagger$,
		\texttt{convnext\_large}$^\dagger$                                                    \\
		\midrule
		Object Detection Models &
		\texttt{retinanet\_resnet50\_fpn\_v2}, \texttt{fcos\_resnet50\_fpn},
		\texttt{fasterrcnn\_resnet50\_fpn\_v2},
		\texttt{ssd300\_vgg16}$^\dagger$, \texttt{ssdlite320\_mobilenet\_v3\_large}$^\dagger$ \\
		\bottomrule
	\end{tabular}

	\vspace{0.1cm}
	{\footnotesize \textsuperscript{$\dagger$} Models used only in cloud cluster evaluation}
\end{table*}

In this section, we describe the experimental setup used to evaluate {\sysname}.

\paragraph{Deployment Infrastructure.}
We conduct our experiments using two distinct deployment types: a cluster of
larger \glspl{gpu} and a cluster of edge \glspl{gpu}. The first consists of
3$\times$ machines equipped with 4$\times$ \textit{Nvidia A100-SXM4-40GB} each,
giving a total of 12 \glspl{gpu}. The second consists of 12$\times$
\textit{Nvidia Jetson AGX Xavier} \glspl{gpu} (referred to as Jetson Xavier for
brevity), also giving a total of 12 \glspl{gpu}. Each GPU is assigned to a
docker to form a server, resulting in 12 servers for each deployment. In
addition, we use 2$\times$ \textit{HPE Proliant DL360 Gen10+}. One machine acts
as the client, which issues inference queries to the system. The other acts as
the controller, which receives these queries and is responsible for scheduling
and forwarding them to the worker servers (each backed by a GPU). This
separation mirrors a typical inference serving setup, where the client generates
requests and the controller orchestrates their distribution to maximize goodput
and efficiency. The full specifications are presented in
\Cref{tab:serve_config}.

\paragraph{Baselines.} We compare {\sysname} against two state-of-the-art
systems: Usher~\cite{shubha2024usher} and INFaaS~\cite{francisco2021infaas}.
Usher is, to the best of our knowledge, the most representative recent work
addressing interference-aware co-location and placement. INFaaS represents a
widely adopted approach to dynamic inference serving. It consists of two
components: a model variant selection module and an auto-scaling module. Since
our work focuses on placement over a pre-specified set of models, we disabled
variant selection; this does not fundamentally alter INFaaS's behavior, as in
its default mode that component treats accuracy as a constraint and selects the
least resource-intensive model that satisfies the query. Other candidates (\eg,
Shepherd~\cite{zhang2023shepherd}, GPUlet~\cite{choi2022serving},
AlpaServe~\cite{li2023alpaserve}) showed radically lower performance than Usher
in its own evaluation, making a direct comparison against Usher sufficient.
Further, systems such as Proteus~\cite{ahmad2024proteus} were excluded because
they target problems orthogonal to ours, such as accuracy auto-scaling.

In all the experiments, we set $\lambda=0.5$ for {\sysname}, \ie, less than the half of performance lost is tolerated in the placement of new models, as described in~\Cref{sec:placement}.

\paragraph{Workloads.} We evaluate our system and baseline methods using both
synthetic and real-world workloads. For the real workload, similar to previous
work we adopt the Twitter trace 2020 dataset~\cite{twitterStreamTrace2020}, as
it is particularly suitable for modeling inference services, as tweets are
commonly subjected to DNN processing before
publication~\cite{francisco2021infaas,ahmad2024proteus}. Since the trace is
aggregated at a coarse temporal granularity of one second, we apply a Poisson
process to model intra-second arrival times and use a Zipf distribution to
distribute queries among models, in line with the established methodology
in~\cite{francisco2021infaas,ahmad2024proteus}. For synthetic workloads, we
generated request rates using a Gaussian process parameterized by the desired
throughput. Model allocation followed a Zipf distribution with an exponent
$\alpha=1.8$, reflecting disparate throughputs between models such as AlexNet
and SSD and producing asymmetric allocations consistent with real-world
imbalances. Finally, note that we also considered the Microsoft Azure Functions
dataset~\cite{shubha2024usher}, but observed similar arrival patterns and
identical experimental trends to the Twitter trace; we therefore prioritized one
real-world dataset complemented by the more controllable synthetic workload.

\paragraph{Models.} To ensure our evaluation captures a broad spectrum of
inference behavior, we selected a diverse and representative set of DNN models.
These include both high-performance classification architectures and widely
adopted object detection frameworks, enabling us to rigorously assess system
behavior under varied computational and latency profiles. The full list of
models is summarized in~\Cref{tab:dnn-models}, reflecting the breadth and
relevance of our evaluation design. Our focus on video analytics models is
motivated by their strict latency requirements, which make interference-aware
placement particularly impactful. Newer LLM-based architectures could in
principle be considered, but they introduce fundamentally different memory
footprints and performance characteristics—such as KV-cache growth and
token-level scheduling—that are incompatible with our edge
setup~\cite{shubha2026adagen}. That said, since LLMs are built on the same
fundamental DNN building blocks, our interference-aware approach could be
extended to them as well. We leave this to future work.

\paragraph{Evaluation Metrics.}
To evaluate the effectiveness of each DNN deployment strategy, the assessment
focused on two categories of metrics: performance metrics and resource metrics.
Performance metrics include SLO violations and goodput. SLO violations serve as
our primary metric, as they directly capture application-level experience: under
a fixed load, rising violations signal that throughput falls short of demand,
and a k\% violation rate is equivalent to a Pk latency exceeding the SLO
threshold. Goodput quantifies the rate of completed requests; note that goodput
curves tend to converge across systems near GPU saturation, and comparisons can
be skewed by heterogeneous models with vastly different peak throughputs (\eg,
AlexNet at $\sim$2000~QPS vs.\ SSD at $\sim$15~QPS), which is why SLO violations
serve as a more robust indicator of tail performance. In our evaluation,
throughput denotes the workload intensity expressed as the submission rate
(queries per second).

Resource metrics include GPU compute utilization, defined as the percentage of
time the GPU's compute engines are actively executing kernels, and memory
utilization, defined as the fraction of GPU memory capacity in use. Together,
these metrics provide a balanced view of service efficiency and hardware usage
under varying workload and batch processing conditions.

%% file: sections/evaluation.tex
\section{Evaluation}\label{sec:evaluation}

This section evaluates {\sysname}'s performance against two state-of-the-art
baselines: INFaaS~\cite{francisco2021infaas} and Usher~\cite{shubha2024usher}.
We examine \sysname's behavior across both cloud-based GPU clusters and edge
deployments using Jetson Xavier devices. Our evaluation aims to answer the
following research questions:

\vspace{0.1cm}\noindent\ding{202} {\bf How does {\sysname} perform compared to existing
	systems in cloud environments?} We demonstrate that {\sysname} reduces SLO
violations by 3$\times$ compared to INFaaS and 2$\times$ compared to Usher while
maintaining comparable or better goodput.

\vspace{0.1cm}\noindent\ding{203} {\bf Can {\sysname} effectively operate on
	resource-constrained edge devices?} We show that {\sysname} achieves similar
	or superior performance on Jetson Xavier devices, keeping SLO violations
	significantly lower than competing baselines despite tight resource
	constraints.

\vspace{0.1cm}\noindent\ding{204} {\bf How does batch size affect {\sysname}'s
scheduling performance under high load?} We demonstrate that {\sysname} achieves
optimal performance at moderate batch sizes (16). Larger batch sizes
increase violations without proportional gains, confirming that efficiency stems
from interference-aware scheduling rather than aggressive batching or raw
resource utilization.

\vspace{0.1cm}\noindent\ding{205} {\bf How accurate is {\sysname}'s placement algorithm under
	varying deployment scenarios?} Through randomized experiments, we find that
	{\sysname} achieves near-optimal placement accuracy in approximately 90\% of
trials, clearly outperforming Usher under identical conditions.

\vspace{0.1cm}\noindent\ding{206} {\bf What is the overhead of {\sysname}'s profiling and
	orchestration mechanisms?} We analyze the computational cost and time
requirements of {\sysname}'s offline profiling and online decision-making
processes, showing negligible impact on overall system performance.


\begin{figure}[t!]
	\centering
	\begin{subfigure}[b]{0.48\linewidth}
		\centering
		\includegraphics[width=\linewidth]{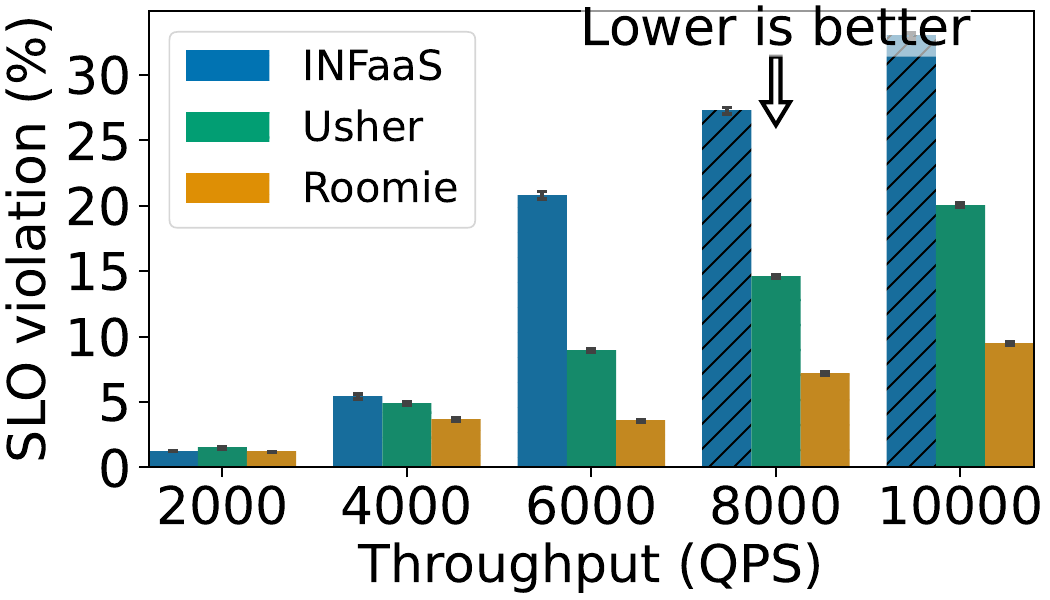}
		\subcaption{SLO violation.}
	\end{subfigure}
	\begin{subfigure}[b]{0.48\linewidth}
		\centering
		\includegraphics[width=\linewidth]{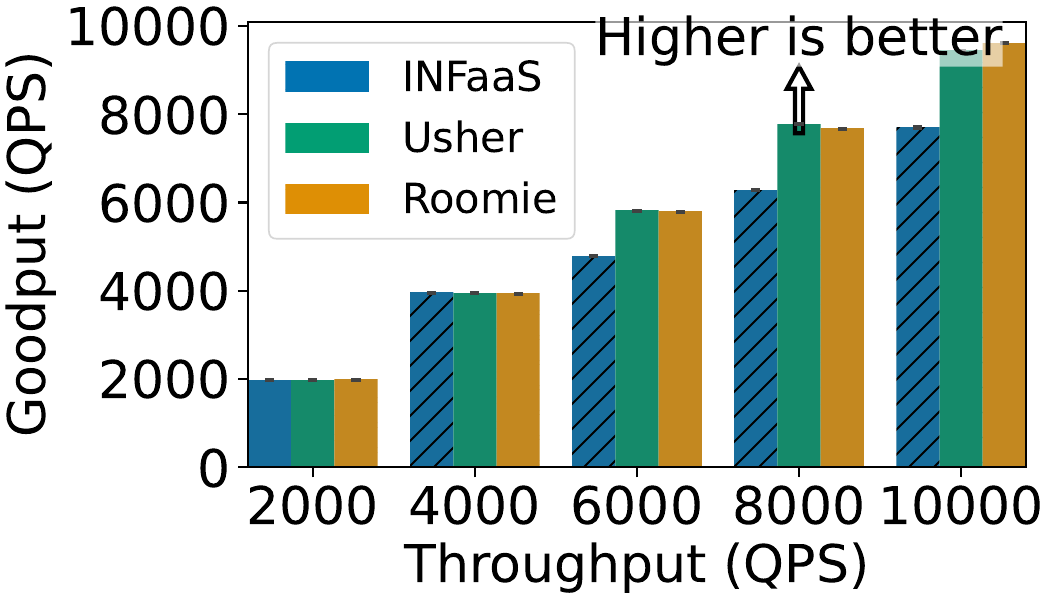}
		\subcaption{Goodput.}
	\end{subfigure}
	\caption{Performance evaluation on the Twitter dataset shows how SLO violations evolve with increasing workload; {\sysname} sustains violations below 10\% under high load, outperforming INFaaS and Usher while preserving goodput.}
	\label{fig:Nvidia/twitter}
\end{figure}

\begin{figure}[t!]
	\centering
	\begin{subfigure}[b]{0.48\linewidth}
		\centering
		\includegraphics[width=\linewidth]{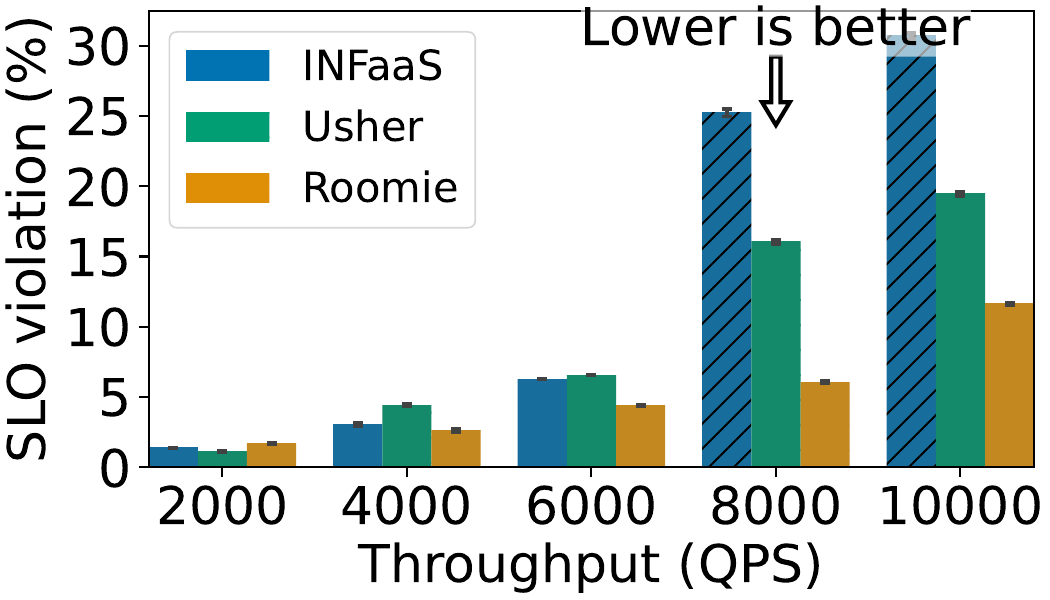}
		\subcaption{SLO violation.}
	\end{subfigure}
	\begin{subfigure}[b]{0.48\linewidth}
		\centering
		\includegraphics[width=\linewidth]{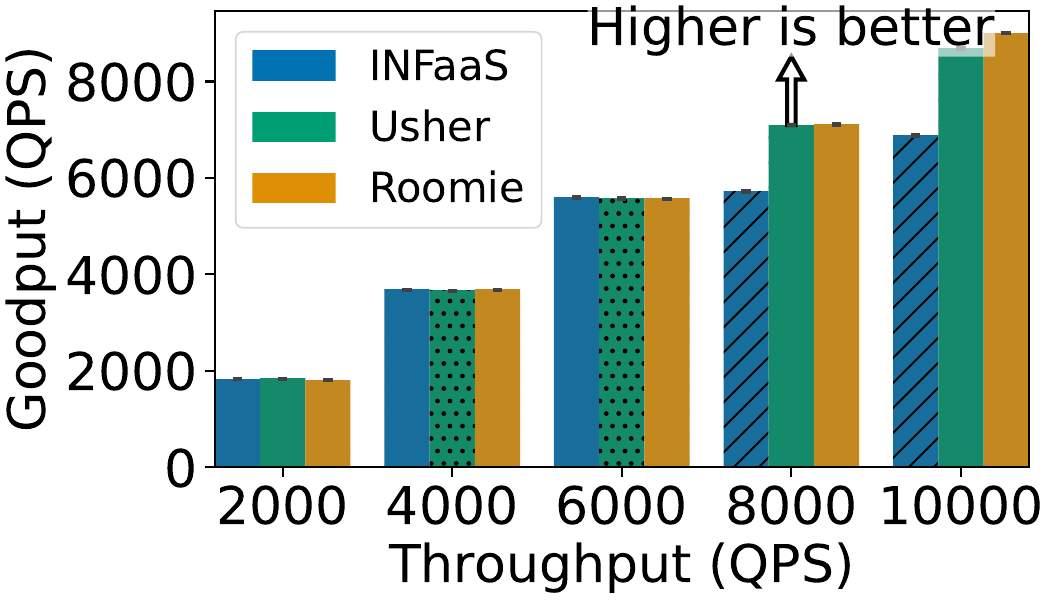}
		\subcaption{Goodput.}
	\end{subfigure}
	\caption{Evaluation with synthetic workloads illustrates system behavior under controlled stress; {\sysname} maintains violations near 6\% at saturation, more than 4$\times$ lower than INFaaS and 2$\times$ lower than Usher, while sustaining goodput close to the offered load.}
	\label{fig:Nvidia/synthetic}
\end{figure}

\subsection{Performance Evaluation of Cloud-Based GPU Cluster Solutions}

To assess the effectiveness of our proposed deployment strategy, we conduct a comprehensive evaluation using a cloud-based GPU cluster comprising 12 \glspl{gpu} and all DNN models detailed in~\Cref{tab:dnn-models}. The experiments are performed using two distinct datasets: real-world Twitter data and synthetically generated data.

\Cref{fig:Nvidia/twitter} shows the performance results obtained from the Twitter dataset. At low workload levels, all approaches behave similarly, with negligible differences in violation rates and goodput. As the workload increases, however, disparities emerge. {\sysname} consistently sustains lower violation rates, achieving less than 10\% even under high load, while INFaaS exceeds 30\% and Usher reaches around 20\%. Goodput remains close to the offered load across all approaches, but {\sysname} maintains slightly higher values, confirming its ability to preserve goodput while reducing violations.

These differences stem directly from how colocation is performed. INFaaS replicates models across workers without interference awareness, which leads to overscaling of heavy detector networks such as SSD, FCOS, and FasterRCNN. When multiple replicas of these models are placed together without awareness, interference between their kernels produces violation rates above 70\% for detectors and more than 60\% for Densenet. Usher attempts to balance compute-bound and memory-bound workloads, but its multiplexing heuristics overlook temporal overlap. As a result, models like MaxViT experience violations around 44\%, and detectors remain unstable with violations between 45--78\%. {\sysname} avoids these drawbacks by distributing heavy models alongside lighter ones whose kernel timelines complement each other. For example, pairing GoogLeNet with SSD allows GoogLeNet to maintain violations below 1\%, while SSD itself remains stressed but at a reduced level. Densenet, which collapses under INFaaS, records only 4.5\% violations under {\sysname}. These placement decisions explain why {\sysname} sustains responsiveness while the baselines degrade.

\Cref{fig:Nvidia/synthetic} illustrates the evaluation conducted with synthetic workloads. The performance trends closely mirror those observed with the Twitter dataset. At high workload levels, {\sysname} maintains violation rates near 6\%, compared to more than 25\% for INFaaS and 16\% for Usher. This corresponds to a reduction of more than 4$\times$ relative to INFaaS and 2$\times$ relative to Usher. Under high workload conditions, the offered goodput is already near the system's maximum capacity for all strategies, but {\sysname} sustains higher goodput than both baselines. Here again, the explanation lies in colocation: INFaaS overscales detectors, Usher misplaces models with overlapping kernels, while {\sysname} minimizes interference by pairing workloads that do not collide in time.

Overall, across both datasets, {\sysname} demonstrates robust performance under varying workload conditions, consistently achieving lower violation rates and higher goodput than competing approaches. This robustness stems from clear factors: replication without interference awareness amplifies contention, multiplexing heuristics that pair compute-heavy with memory-heavy models fail to capture temporal dynamics, and interference-aware colocation prevents collapse by placing DNN models intelligently.

\begin{figure}[t!]
	\centering
	\begin{subfigure}[b]{0.48\linewidth}
		\centering
		\includegraphics[width=\linewidth]{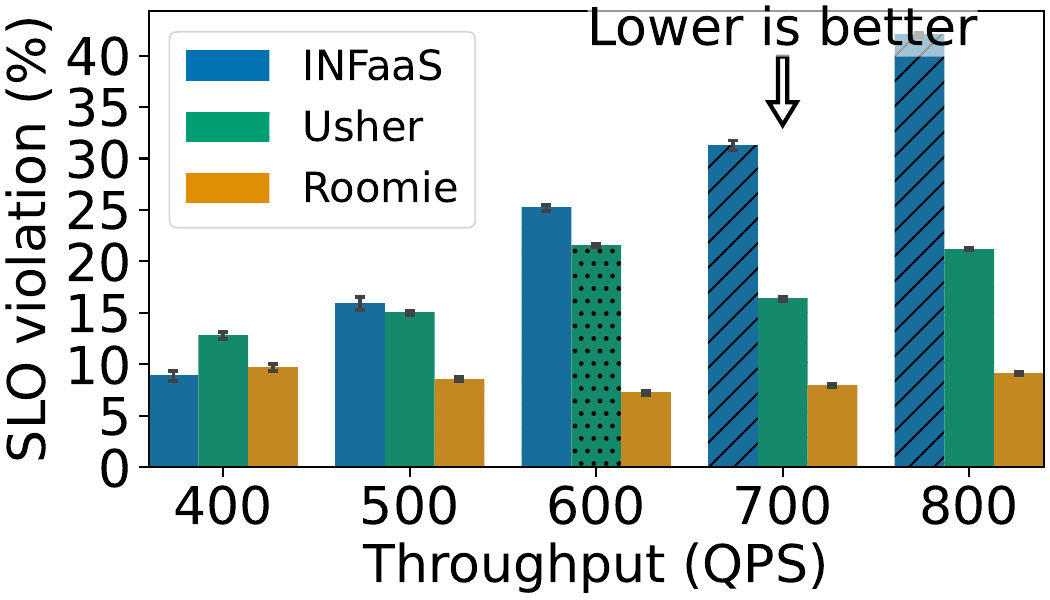}
		\subcaption{SLO violation.}
	\end{subfigure}
	\begin{subfigure}[b]{0.48\linewidth}
		\centering
		\includegraphics[width=\linewidth]{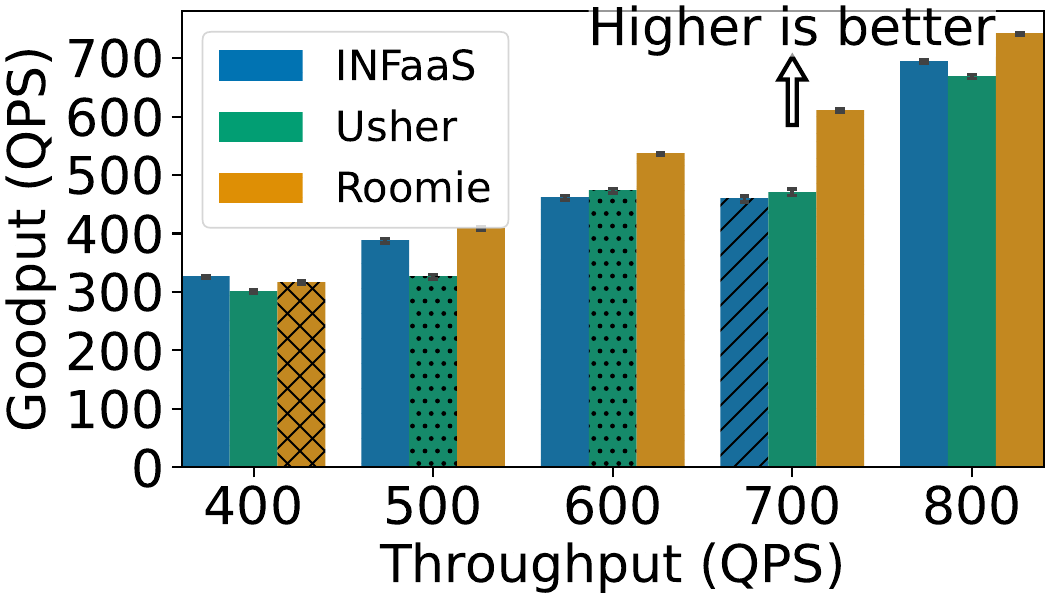}
		\subcaption{Goodput.}
	\end{subfigure}
	\caption{Edge-based evaluation using the Twitter dataset demonstrates the impact of resource constraints; {\sysname} keeps violations near 9\% under high load, compared to over 42\% for INFaaS and 21\% for Usher, while sustaining higher goodput.}
	\label{fig:JetsonNano/twitter}
\end{figure}

\begin{figure}[t!]
	\centering
	\begin{subfigure}[b]{0.48\linewidth}
		\centering
		\includegraphics[width=\linewidth]{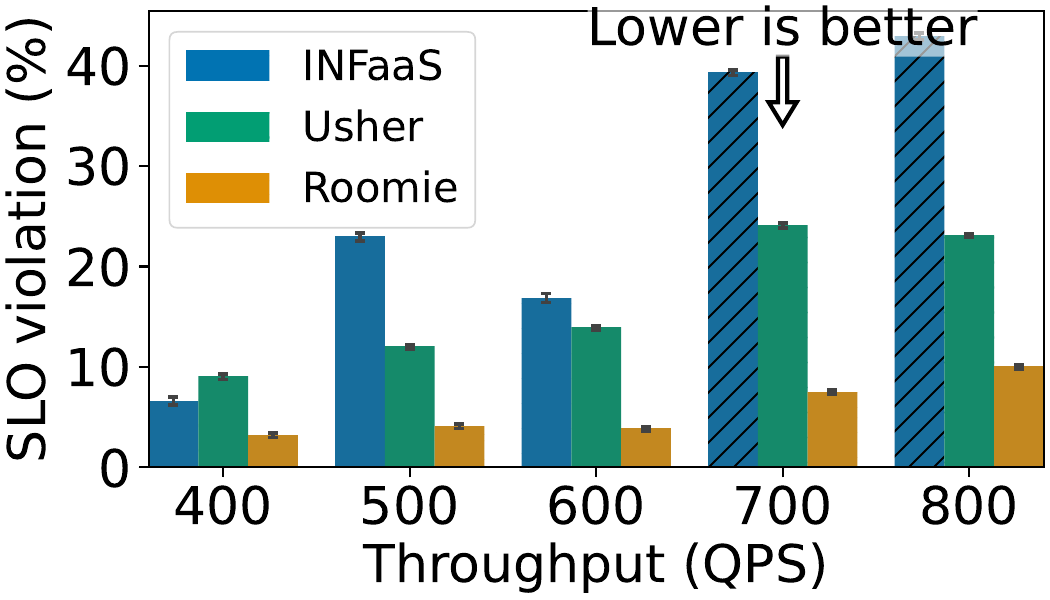}
		\subcaption{SLO violation.}
	\end{subfigure}
	\begin{subfigure}[b]{0.48\linewidth}
		\centering
		\includegraphics[width=\linewidth]{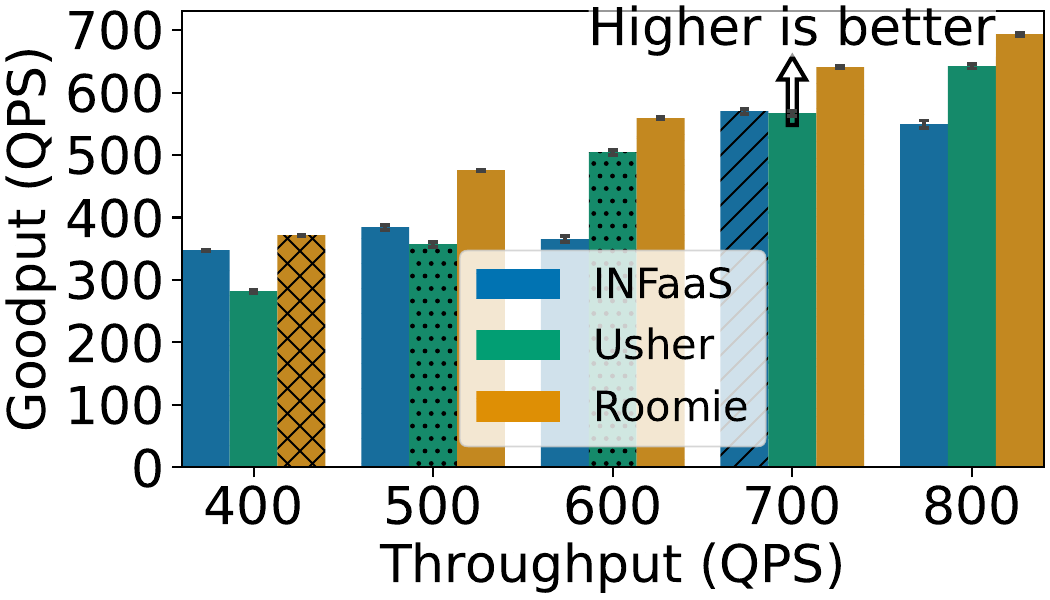}
		\subcaption{Goodput.}
	\end{subfigure}
	\caption{Evaluation with synthetic workloads on Jetson Xavier devices highlights robustness under saturation; {\sysname} maintains violations close to 10\%, reducing them by factors of four or more relative to INFaaS and outperforming Usher in both violation rate and goodput.}
	\label{fig:JetsonNano/synthetic}
\end{figure}

\subsection{Performance Evaluation on Edge Devices Using Jetson Xavier \glspl{gpu}}

To further validate our solution, we conduct a second set of experiments using a cluster of 12 Jetson Xavier \glspl{gpu}, representative of resource-constrained edge computing environments. As in the cloud-based evaluation, we deploy 12 models (from~\Cref{tab:dnn-models}), which correspond to the \glspl{dnn} not marked with $^\dagger$ in the table, and tested performance using real-world Twitter data and synthetically generated data while gradually increasing workload intensity.

The results of the Twitter workload on Jetson Xavier devices appear in~\Cref{fig:JetsonNano/twitter}. At low traffic levels, the three approaches deliver comparable performance, with only minor differences in violation rates and goodput. As the workload grows, however, {\sysname} begins to separate itself from the baselines. At moderate intensity, it sustains lower violation rates while maintaining goodput close to the offered load. Under high workload conditions, the contrast becomes pronounced: {\sysname} holds violations near 9\%, whereas INFaaS rises above 40\% and Usher remains above 20\%. Goodput measurements confirm this advantage, with {\sysname} consistently achieving higher goodput than both competitors.

\begin{figure}[t!]
	\centering
	\begin{subfigure}[b]{0.48\linewidth}
		\centering
		\includegraphics[width=\linewidth]{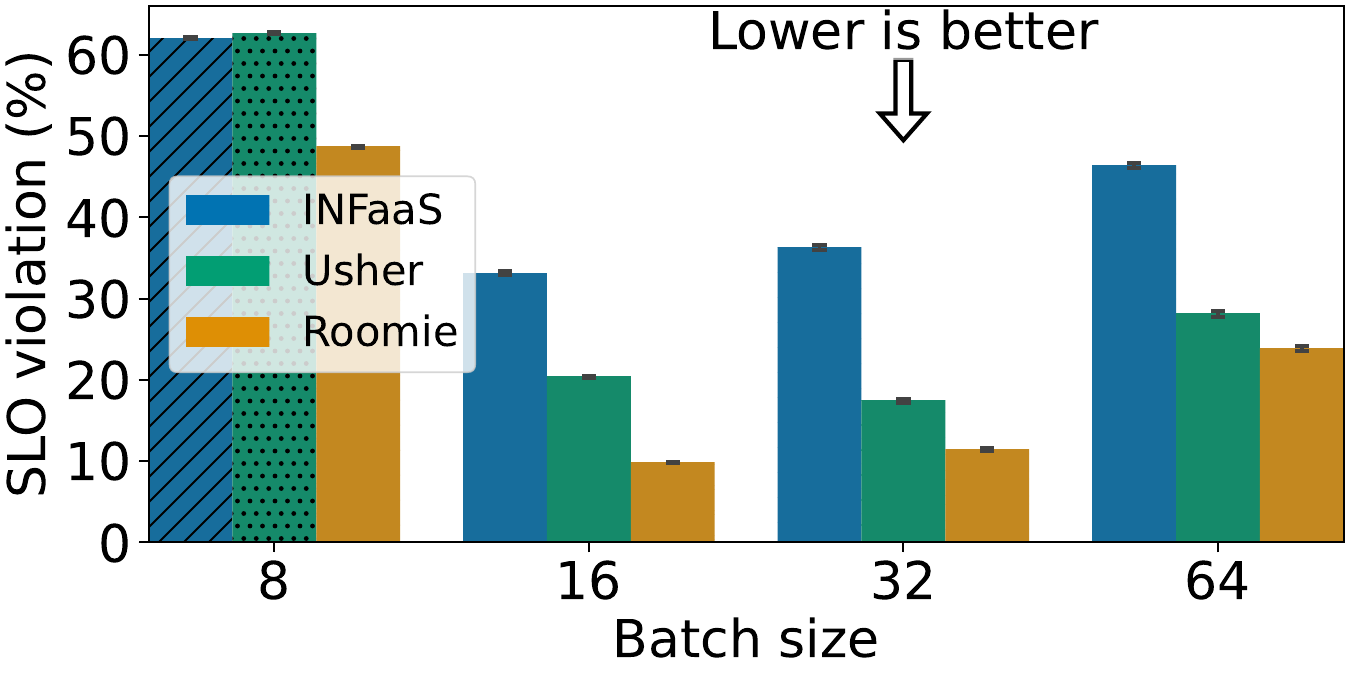}
		\subcaption{SLO violation.}
	\end{subfigure}
	\begin{subfigure}[b]{0.48\linewidth}
		\centering
		\includegraphics[width=\linewidth]{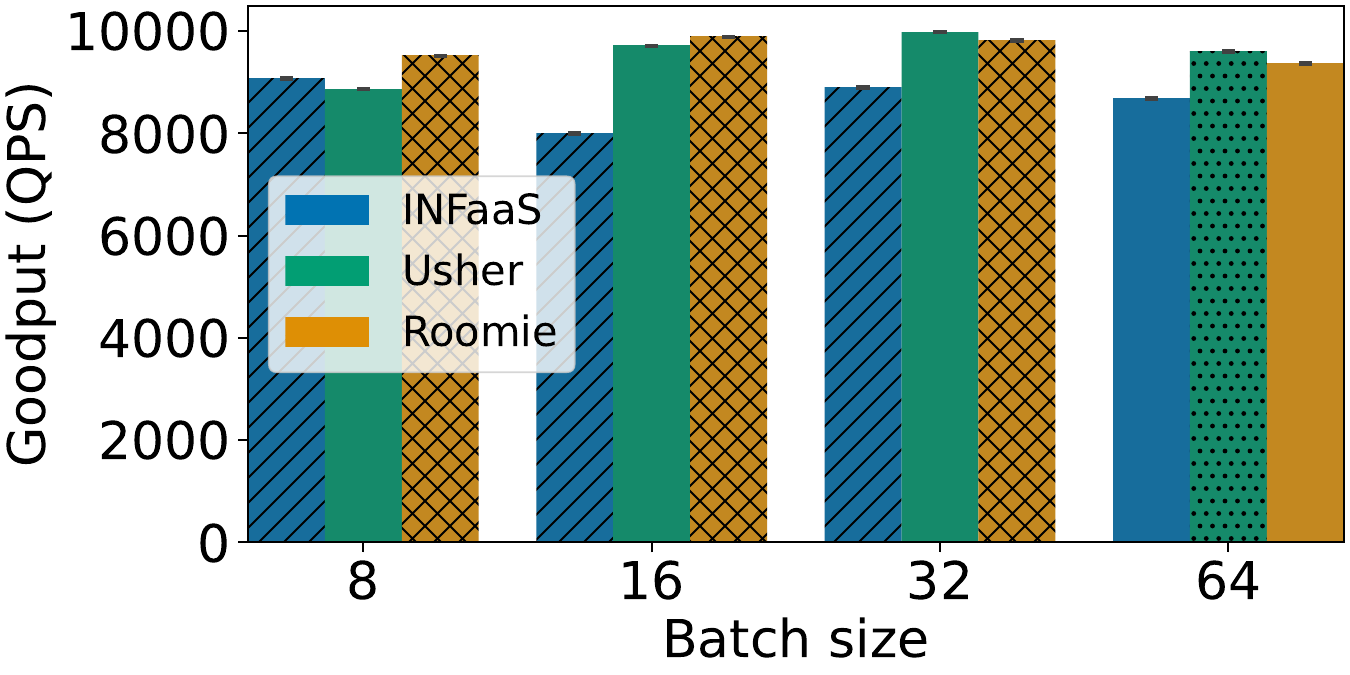}
		\subcaption{Goodput.}
	\end{subfigure}
	\caption{Impact of batch size on SLO violations and goodput under high workload (10K QPS); while goodput remains nearly unchanged, larger batches increase violations across all approaches, with {\sysname} consistently maintaining lower rates—especially at moderate batch sizes.}
	\label{fig:BatchSize/Nvidia_perf_metrics}
\end{figure}

These results complement the conclusions drawn from the evaluation of cloud-based clusters. Replication without interference consideration causes INFaaS to collapse under saturation, while Usher's strategy of pairing compute-heavy with memory-heavy models offers partial improvements but fails to capture kernel overlap. On Jetson devices, the same mechanisms are visible but are magnified by tighter resource constraints. INFaaS again overscales detectors, with Retinanet exceeding 73\% violations, while even lightweight models such as AlexNet (23\%), GoogLeNet (34\%), and MobileNet (35\%) degrade when colocated with heavy workloads. Usher continues to misplace heterogeneous models, producing moderate violations for classifiers such as AlexNet (21\%) and Densenet (12\%), while FCOS and MaxViT remain unstable at 29\% and 27\%. By contrast, {\sysname} places heavy models alongside lighter ones whose kernel timelines complement each other, reducing contention for most workloads: Densenet's violations fall to 5\%, MobileNet to 4\%, Shufflenet to 6\%, and AlexNet to 6\%. Detectors such as FasterRCNN and FCOS remain stable under {\sysname}, both near 1--2\%, though Retinanet continues to be challenging at 62\%. These examples illustrate how interference-aware colocation prevents collapse and sustains responsiveness under stress, even though certain detector models remain difficult to stabilize.

The synthetic dataset evaluation on Jetson Xavier \glspl{gpu} produces results consistent with those observed on the Twitter dataset, as shown in~\Cref{fig:JetsonNano/synthetic}. Under high load, {\sysname} maintains violation rates close to 10\%, compared to more than 40\% for INFaaS and 20\% for Usher. At these traffic levels, the incoming query rate drives the system near saturation across all strategies, yet {\sysname} converts a significantly larger share of requests into successful completions—achieving up to 1.5$\times$ higher goodput than competing baselines. Here again, the explanation lies in colocation: INFaaS overscales detectors, Usher misplaces models with overlapping kernels, while {\sysname} minimizes interference by pairing workloads that do not collide in time.

Taken together, the evaluation on edge devices confirms and strengthens the earlier cluster-based conclusion. Replication without interference awareness leads to collapse under saturation, multiplexing compute-heavy with memory-heavy models provide partial improvements but fails to capture temporal dynamics, and interference-aware colocation is essential for sustaining responsiveness and goodput. The sharper contrasts observed on Jetson Xavier devices highlight that the benefits of {\sysname} are not limited to large clusters but extend to resource-constrained environments, demonstrating its generality and robustness across deployment contexts.

\begin{figure}[t!]
	\centering
	\begin{subfigure}[b]{0.48\linewidth}
		\centering
		\includegraphics[width=\linewidth]{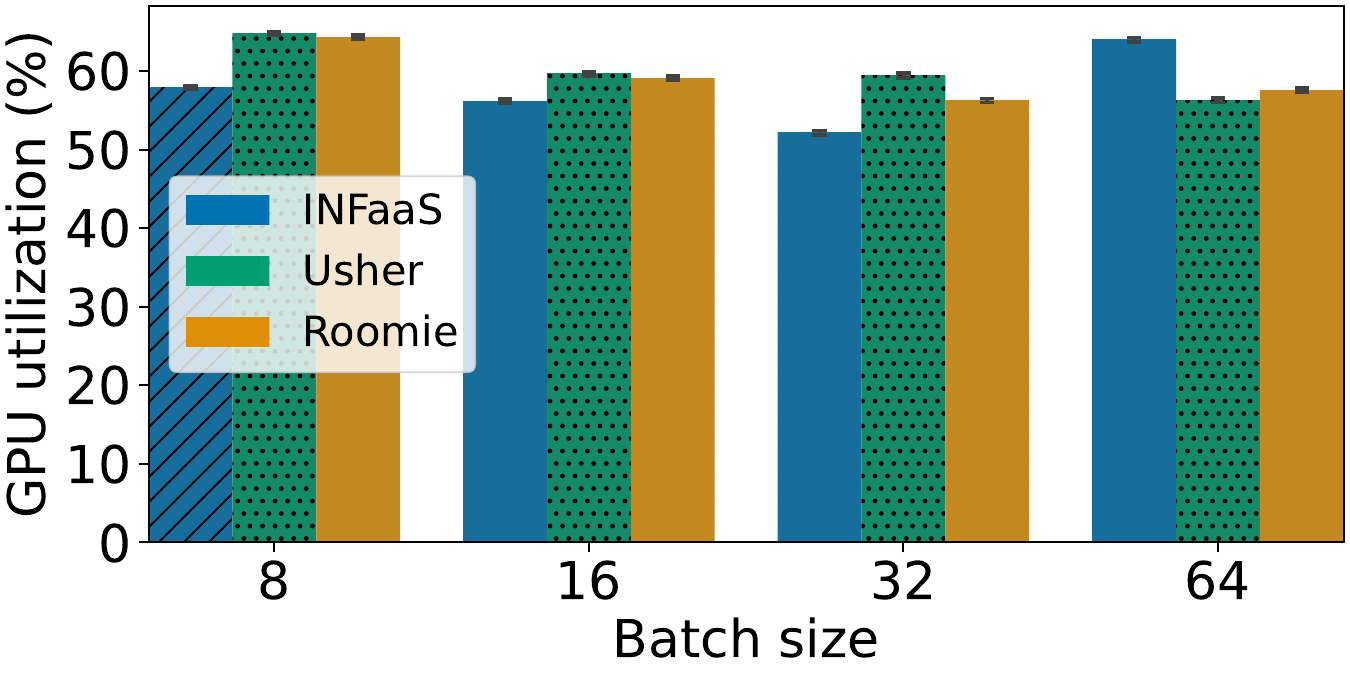}
		\subcaption{GPU core utilization.}
	\end{subfigure}
	\begin{subfigure}[b]{0.48\linewidth}
		\centering
		\includegraphics[width=\linewidth]{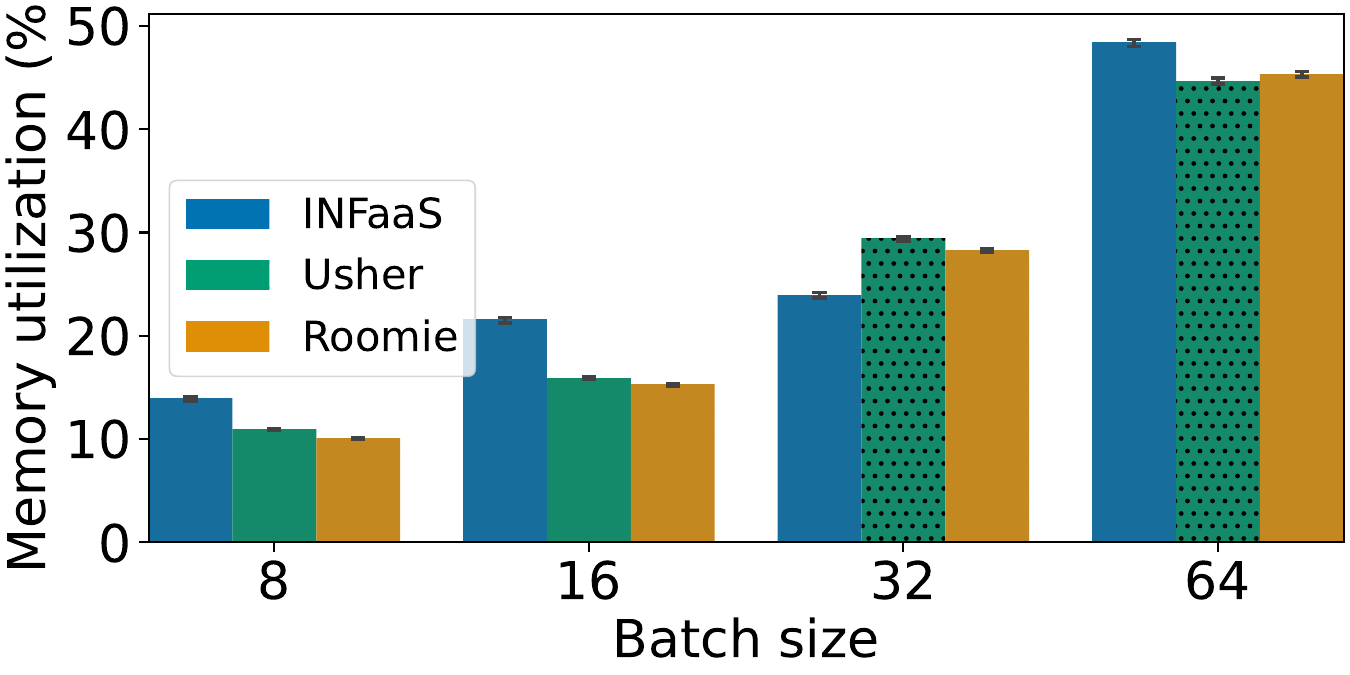}
		\subcaption{Memory utilization.}
	\end{subfigure}
	\caption{Impact of batch size on GPU core utilization and memory utilization under high workload (10K QPS) ; GPU core utilization shows little variation due to CUDA kernel execution patterns, while memory usage increases with larger batches but exhibits only modest differences across approaches.}
	\label{fig:BatchSize/Nvidia_resource_metrics}
\end{figure}

\subsection{Impact of batch size}

We examine the effect of batch size on scheduling performance using the cloud-based GPU cluster with 12 \glspl{gpu}. The experiment uses the Twitter dataset under high workload conditions (10K QPS) and varies batch size from 8 to 64. To capture the impact of batching under stress, we analyze performance metrics (SLO violations and goodput) alongside resource metrics (GPU core and memory utilization), providing a comprehensive view of how batching influences efficiency and service quality.

The performance metrics in~\Cref{fig:BatchSize/Nvidia_perf_metrics} reveal decisive effects. At batch size 8, all approaches exhibit high violation rates, though {\sysname} consistently maintains lower violations than INFaaS and Usher. Increasing to batch size 16 produces a pronounced improvement for {\sysname}, with violations dropping to a small fraction of those at batch size 8 and remaining lower than both baselines, while goodput stays nearly unchanged. At larger batches, violations rise again across all solutions relative to the improvement observed at 16, indicating that aggressive batching increases responsiveness penalties even though goodput changes only marginally. In summary, when goodput is already high at smaller batches, further increasing batch size is not necessarily advantageous; it tends to worsen violations without yielding proportional goodput gains.

\begin{figure}[t!]
	\centering
	\includegraphics[width=.7\linewidth]{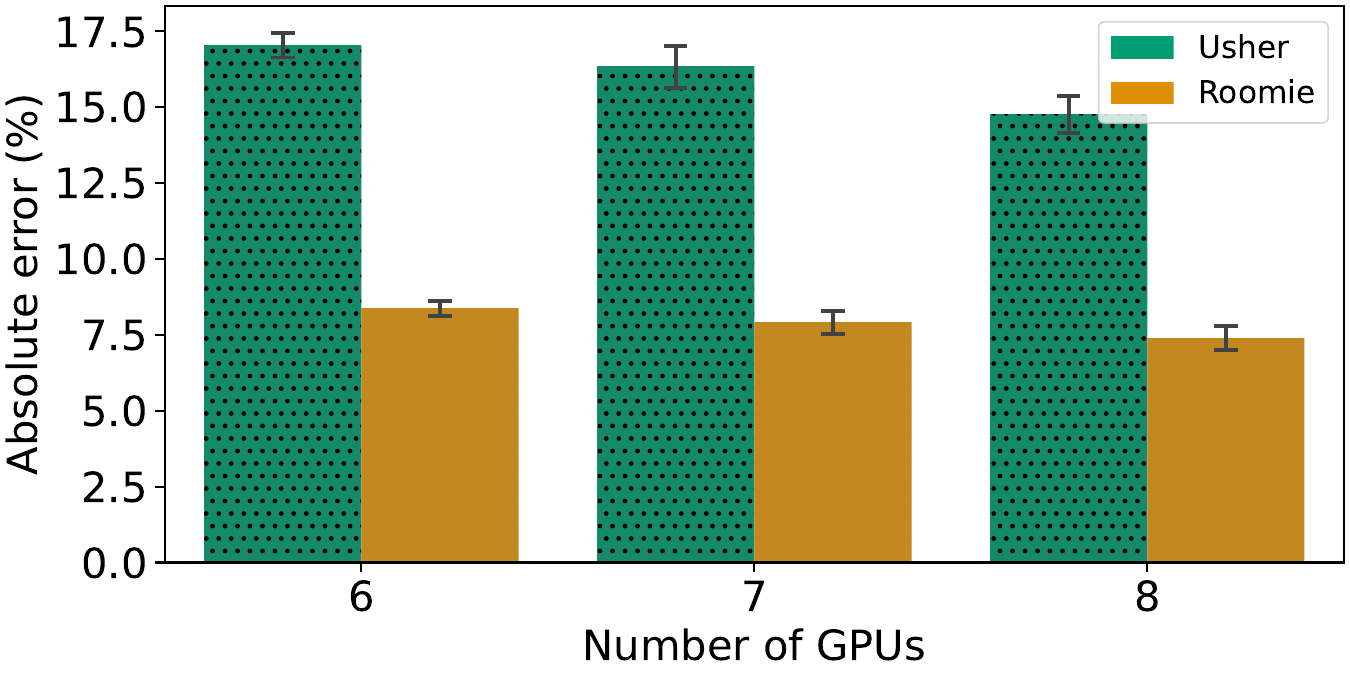}
	\caption{{\sysname} maintains deployment error within 7--8\% of the optimal and outperforms Usher in nearly 90\% of evaluated scenarios.}
	\label{fig:performance_gap}
\end{figure}

The resource metrics in~\Cref{fig:BatchSize/Nvidia_resource_metrics} show that goodput-oriented batching does not materially differentiate the approaches in this setup. GPU core utilization remains stable across batch sizes for all strategies, reflecting the behavior of CUDA kernels, which maximize occupancy regardless of workload configuration. Memory utilization increases with larger batches, but the differences between approaches are modest and do not translate into meaningful changes in service quality.

The experiments show that responsiveness in multi-tenant inference serving inference is not determined by raw execution speed or aggregate resource utilization, but by how workloads are colocated and scheduled on GPU resources. {\sysname} consistently reduces SLO violations while sustaining comparable goodput, demonstrating that efficiency stems from interference-aware scheduling rather than attempts to drive GPU or memory usage higher. This is emphasized by the fact that GPU core utilization remains essentially unchanged across approaches and batch sizes, and memory utilization differences are modest, confirming that resource metrics do not account for the observed performance gap. The decisive factor is whether planning strategies recognize and mitigate interference between DNN kernels. {\sysname} derives its advantage precisely from adopting a colocation approach that takes this interference into account, allowing it to maintain its responsiveness in high-demand scenarios while preserving its efficiency.

\begin{table}[t!]
\centering
\small
\setlength{\tabcolsep}{10pt}
\caption{Total decision latency (ms) by number of deployed DNNs and GPUs.}
\label{tab:decision_latency}
\begin{tabular}{@{}cccc@{}}
\toprule
\multirow{2}{*}{\textbf{\# Models}} & \multicolumn{3}{c}{\textbf{Total decision latency (ms)}} \\
\cmidrule(lr){2-4}
 & \textbf{2 GPUs} & \textbf{5 GPUs} & \textbf{10 GPUs} \\
\midrule
5  & 184.8          & 85.5            & 110.4 \\
10 & ---            & 259.9           & 127.1 \\
15 & ---            & 776.9           & 378.4 \\
\bottomrule
\end{tabular}
\end{table}


\subsection{Deployment Accuracy Evaluation against Optimal Strategies}

This section investigates the effectiveness of {\sysname} for the deployment of \glspl{dnn} across a varying number of \glspl{gpu}, specifically between six and eight. For each configuration, the number of \glspl{dnn} to be deployed is randomly selected from a range of 2--3$\times$ the number of \glspl{gpu}, guided by predefined options outlined in~\Cref{tab:dnn-models}. More than 1500 randomized evaluations are conducted to ensure comprehensive coverage of deployment scenarios. In each evaluation, all feasible deployment permutations are thoroughly assessed to determine the configuration that results in the minimal average performance drop, defined as the optimal baseline. Both {\sysname} and Usher are then applied to the same scenarios, and their absolute errors relative to the optimal are recorded. The results are summarized in~\Cref{fig:performance_gap}, which illustrates the average performance gap across configurations.

The comparative analysis highlights {\sysname}'s consistent superiority in deployment accuracy across all concurrency levels. Its success in nearly 90\% of randomized trials reflects a design that is not only structurally aware but also resilient to the practical limitations of kernel-level modeling. Unlike Usher, which applies static heuristics that overlook the dynamic nature of interference, {\sysname} adapts to the complexities introduced by concurrent execution. Crucially, the residual error observed in {\sysname}'s deployments stems not from heuristic misalignment, but from the inherent challenges of profiling-based estimation. Tools such as Nsight-Compute, while indispensable for capturing fine-grained kernel behavior, introduce latency and measurement distortion that complicate performance inference. {\sysname}'s strategy, based on the analysis of isolated traces and representative overlap simulation, effectively manages these distortions without resorting to exhaustive enumeration. Moreover, as concurrency increases, {\sysname} demonstrates robustness in the face of combinatorial explosion, where kernel alignment across models creates an exponentially growing space of interference scenarios. That {\sysname} maintains bounded error under these conditions affirms its capacity to balance fidelity with scalability, offering a principled alternative to heuristics that fail to account for architectural nuance.

\subsection{Latency Evaluation for DNN Colocation}

This evaluation setup is designed to assess how GPU count and the number of
deployed DNNs influence the computational effort required to analyze colocation
scenarios. In each experiment, DNN models are randomly sampled from the set of
architectures in~\Cref{tab:dnn-models} and deployed across 2, 5, or 10 GPUs. For
every arrival, the interference-aware placement algorithm evaluates all
candidate GPUs before finalizing a decision. We measure the total decision
latency, defined as the cumulative time to deploy all DNNs in a run.
Table~\ref{tab:decision_latency} reports the aggregated ranges for total
decision latency across repeated runs. We do not report numbers for 10 and 15
models on 2 GPUs, as beyond 3 models per GPU the available GPU memory is
exceeded or latency violates the SLO by a wide margin.

The results show that decision latency remains low when the number of GPUs is
greater than or equal to the number of models to be deployed, or when the number
of DNNs is low. For example, with 5 DNNs, the total latency is approximately 85
ms on 5 GPUs and 110 ms on 10 GPUs, while 2 GPUs require more time (~185 ms). As
deployments scale up, total latency increases, reaching approximately 260 ms for
10 DNNs on 5 GPUs and nearly 780 ms for 15 DNNs on 5 GPUs. Even in the most
demanding case of 15 DNNs on 10 GPUs, the totals remain below 400 ms, showing
that the system scales without major time-related issues. Lightweight models or
architectures such as AlexNet or Squeezenet consistently produce negligible
latencies, while SSD variants dominate the tail and drive cumulative values up
to several hundred or even several thousand milliseconds, due the high number of
kernels to evaluate. When SSD models are excluded, the totals drop sharply. For
example, 15 DNNs on 5 GPUs drop to around 250 ms and on 10 GPUs to around 150
ms, confirming that the system can support larger deployments with decision
times well below one second.

%% file: sections/related.tex
\section{Related Work}\label{sec:related}

Scheduling large-scale inference workloads on GPU clusters has become a central
problem as deep learning services proliferate. Unlike training, inference must
simultaneously satisfy latency, accuracy, and cost objectives, which are often
in tension and require specialized scheduling solutions.

\paragraph{Inference Serving Systems.}
Clipper~\cite{2017clipper} and
TensorFlow-Serving~\cite{olston2017tensorflowserving} simplify model deployment
and adapt to traffic by scaling replicas, but neither accounts for interference
for colocated models. Clockwork~\cite{gujarati2020servingDNNlikeclockwork}
achieves predictable performance by executing one inference at a time, at the
cost of underutilized GPUs. Proteus~\cite{ahmad2024proteus} introduces adaptive
batching and dynamic variant selection, but restricts each device to a single
variant, limiting parallel execution. More recently,
AdaGen~\cite{shubha2026adagen} proposes a workload-adaptive scheduler for LLM
inference; its focus on KV-cache and token-level scheduling makes it orthogonal
to our setting of colocating traditional DNN models on heterogeneous GPUs.

\paragraph{Multi-Tenant DNN Inference on Shared GPUs.}
Recent work has explored concurrent execution of multiple models on shared
GPUs. INFaaS~\cite{francisco2021infaas} enables model-less serving by
dynamically selecting variants and reactively scaling workers, but lacks
proactive interference prediction. Colti~\cite{mobin2023colti} improves
throughput by colocating training and inference workloads, while Yu et
al.~\cite{yu2021automated} exploit operator-level independence to schedule
concurrent execution across streams. REEF~\cite{han2022microsecond} and
Miriam~\cite{zhao2023miriam} support kernel preemption and elastic kernels for
priority-based, real-time scheduling. While these systems optimize
post-deployment execution, they overlook initial placement decisions that could
preemptively mitigate interference.

\paragraph{Interference-Aware Inference Serving.}
A growing body of work models interference between concurrently running models
to drive proactive scheduling. Mendoza \etal.~\cite{mendoza2021interference}
predict latency degradation from global buffer and PCIe utilization, but coarse
granularity limits accuracy. Scrooge~\cite{hu2021scrooge} profiles concurrency
thresholds for identical \glspl{dnn}, an approach infeasible for heterogeneous
combinations due to profiling overhead. Abacus~\cite{cui2021Abacus} jointly
schedules operators across models to maintain QoS, but its hardware-agnostic
duration model and reactive execution lead to underutilization.
iGnifer~\cite{xu2023iGniter} characterizes interference from GPU metrics such
as L2 cache usage and core launch counts, though coarse indicators like power
consumption prove less predictive. Usher~\cite{shubha2024usher} refines this
direction by analyzing kernel occupancy and DRAM usage to distinguish compute-
from memory-intensive workloads. Both Usher and iGnifer, however, rely on
Nvidia's \gls{mps}~\cite{nvidiaMPS575} for spatial sharing, limiting their
applicability to edge platforms such as Nvidia Jetson, where MPS is
unsupported.

%% file: sections/conclusion.tex
\section{Conclusion}\label{sec:conclusion}
We presented {\sysname}, an interference-aware serving system for colocated DNN
inference on shared GPUs. {\sysname} couples an offline kernel-level profiling
and interference estimation framework with an online placement algorithm that
assigns models to GPUs under a bounded performance-degradation budget. Evaluated
on both a 12-GPU Nvidia A100 cloud cluster and a 12-device Jetson Xavier edge
deployment, {\sysname} reduces SLO violations by up to 3$\times$ over INFaaS and
2$\times$ over Usher while matching the optimal placement in roughly 90\% of
randomized trials and keeping decision latency below one second for up to
fifteen models on ten GPUs. These results show that modeling \emph{when} kernels
collide in time, rather than which resources they nominally consume, is what
makes interference-aware placement practical across both cloud and edge.